\documentclass[reprint,aip,jcp ,superscriptaddress]{revtex4-2}
\usepackage{amssymb}
\usepackage{amsmath}
\usepackage{graphicx}
\usepackage{colortbl}
\usepackage{xcolor}
\usepackage[caption=false]{subfig}
\usepackage[hidelinks]{hyperref}
\usepackage{comment}
\usepackage[english]{babel}
\usepackage{siunitx}
\usepackage{float}
\usepackage{comment}

\newcommand{\dif}{\text{d}}

\newcommand{\mean}[1]{\left\langle #1 \right\rangle}
\newcommand{\MBR}{\text{MBR}}
\newcommand{\VBR}{\text{VBR}}
\newcommand{\mbr}{\mathcal{B}}
\newcommand{\abs}[1]{\left\vert #1 \right\vert}

\begin{document}
\title{Evaluating non-equilibrium trajectories via mean back relaxation: Dependence on length and time scales}

\author{Gabriel Knotz}
\email[]{g.knotz@theorie.physik.uni-goettingen.de}
\affiliation{Institute for Theoretical Physics, Göttingen, Germany}
\author{Till M. Muenker}
\affiliation{Third Institute of Physics, Göttingen, Germany}
\author{Timo Betz}
\affiliation{Third Institute of Physics, Göttingen, Germany}
\author{Matthias Krüger}
\email[]{matthias.kruger@uni-goettingen.de}
\affiliation{Institute for Theoretical Physics, Göttingen, Germany}

\begin{abstract}
    The mean back relaxation (MBR) relates the value of a stochastic process at three different time points. It has been shown to detect broken detailed balance under certain conditions. For experiments of probe particles in living and passivated cells, MBR was found to be related to the so called effective energy, which quantifies the violation of the fluctuation dissipation theorem. In this manuscript, we discuss the dependence on the length and time parameters that enter MBR, both for cells as well as for a model system, finding qualitative agreement between the two. For the cell data, we extend the phenomenological relation between MBR and effective energy to a larger range of time parameters compared to previous work, allowing to test it in systems with limited resolution.
    We analyze the variance of back relaxation (VBR) in dependence of the mentioned parameters, relevant for the statistical error in  MBR evaluation. For  Gaussian systems, the variance is found analytically in terms of the  mean squared displacement, and we determine its absolute minimum as a function of the length and time parameters. Comparing VBR from cell data to a Gaussian prediction demonstrates a non-Gaussian process. 
     
\end{abstract}
\maketitle
\section{Introduction}   

Over the past decades, studies have shown that the mechanical properties of cells play a key role in many essential cellular functions like migration, proliferation, and development, and often these properties are dysregulated in severe diseases like cancer or Covid-19 \cite{roca-cusachs_quantifying_2017,petridou_multiscale_2017,mohammadi_mechanisms_2018, van_helvert_mechanoreciprocity_2018,Guck2005-qj,Bufi2015-pw,Hurst2021-lp,Kubankova2021-kb}. A plethora of studies has demonstrated this by probing the viscoelastic material properties of cells from the outside 
using techniques like atomic force microscopy \cite{Rother2014-cq}, flow cytometry \cite{Otto2015-gh}, or micropipette aspiration \cite{Hochmuth2000-tw}. However, even though recent studies emphasize that also intracellular mechanics are relevant for many cellular processes \cite{Catala-Castro2025-yc,Muenker2024-aj, Hurst2021-lp}, they have been less extensively studied. One reason for this are the elaborate experimental techniques required to probe these intracellular mechanical properties.

Passive techniques, like particle tracking microrheology \cite{Mason1995-cg},  infer the mechanical properties in form of response functions or shear moduli by measuring fluctuations of the unperturbed system, relying on the  fluctuation dissipation theorem (FDT) \cite{kubo_fluctuation-dissipation_1966, agarwal_fluctuation-dissipation_1972,harada_equality_2005,speck_restoring_2006,baiesi_fluctuations_2009,baiesi_update_2013} 
\begin{align}
    C(\omega)=\frac{2 k_BT \chi^{\prime\prime}(\omega)}{\omega}.
    \label{FDT:classical}
\end{align}
It connects the imaginary  part of the response function $\chi(\omega)$ with the power spectral density $C(\omega)$, with frequency $\omega$, and thermal energy $k_BT$. However, this powerful theorem is only applicable in equilibrium. Out of equilibrium, the relation breaks down, which can be prominently observed in cells \cite{guo_2014, Turlier2016-dg, ahmed_active_2018,stoev_active_2025}. Therefore, intracellular mechanical properties are classically determined by applying external forces to a probe particle and observing the particle motion in response to these forces, i.e., thereby directly measuring the response function $\chi(\omega)$. This requires elaborate experimental techniques like optical or magnetic tweezers\cite{Vos2024-wo,Nishizawa2017-nb,Ebata2023-pi}. 
It has been found that, in cells,  the above relation can be generalized by introduction of a so called effective energy $E_\text{Eff}$ \cite{martin_comparison_2001,mizuno_nonequilibrium_2007,ahmed_active_2018},
\begin{align}
    C(\omega)=\frac{2 E_\text{Eff}(\omega) \chi^{\prime\prime}(\omega)}{\omega}.
    \label{FDT:adapted}
\end{align}
For example, for a probe particle immersed in cells of various types \cite{Muenker2024-aj}, it was  found, using an optical tweezer setup, that  the effective energy follows a power law, i.e., $ \frac{E_\text{Eff}(\omega)}{k_B T} \simeq 1 + \frac{\Omega}{\omega}$, with the frequency prefactor $\Omega$ depending on the cell under investigation \cite{Muenker2024-aj}.

Because detailed balance is considered a sufficient condition for FDT in Eq.~\eqref{FDT:classical} \cite{risken_fokker-planck_1996}, such a finding of $E_\text{Eff}\not=k_BT$ marks the violation of detailed balance, whose inference is generally considered a non-trivial problem \cite{martinez_inferring_2019, battle_2016}. This discussion emphasizes that the quest to obtain mechanical properties of active systems such as cells is intimately connected to the question of broken detailed balance. 

In previous work, we introduced  the so called mean back relaxation (MBR) \cite{muenker_accessing_2024,knotz_mean_2024,ronceray_two_2023}, which correlates particle positions at three different times. More specifically,  MBR quantifies the dependence of future displacements of a probe particle on its past trajectory. In Ref.~\onlinecite{muenker_accessing_2024}, we  observed a linear relationship between the above introduced $E_\text{eff}$ and the long time  limit of MBR. Such relation was also observed in a simple model system \cite{muenker_accessing_2024,john_progress_2025} of a probe particle in a randomly moving trap. In Ref.~\onlinecite{knotz_mean_2024}, we showed that, if the particle position relaxes to a well defined  mean position in a finite amount of time, MBR reaches, for long times, the value of $\frac{1}{2}$ in a system with detailed balance. Under the mentioned conditions,  any deviation of $\frac{1}{2}$ thus marks broken detailed balance \cite{dieball_perspective_2025}. MBR thus appears to have promising properties, which we aim to investigate further in this manuscript.

MBR, introduced in detail below,  depends on two time parameters, $t$ and $\tau$, and one length parameter, $l$ (see Eq.~\eqref{eq:MBR} below). Here we discuss the dependence of MBR on the time parameters in detail, both for cells and for a model system. We expand the model introduced in Ref.~\onlinecite{muenker_accessing_2024} by one more degree of freedom. This modification of the model incorporates memory and reproduces the characteristic non-monotonic behavior of MBR as a function of time $t$, consistent with observations in cellular data \cite{muenker_accessing_2024}. The amendment further introduces a dependence on the so called conditioning time $\tau$, which is absent in the simpler model of Ref.~\onlinecite{muenker_accessing_2024}. MBR of the so amended model is discussed in detail, and shown to qualitatively agree with cell data as concerns the dependence on the two mentioned times. This also allows to analyze and discuss the relation between MBR and effective energy for the larger range of time parameters compared to Ref.~\onlinecite{muenker_accessing_2024}. This is, e.g., relevant for experimental setups with lesser temporal resolution.

We analyze the {\it variance} of  back relaxation (VBR), relevant for the statistical error of MBR. We provide an analytical solution for VBR for any Gaussian process, finding that it, in contrast to MBR, depends on the parameter $l$ (and also on $t$ and $\tau$). On the level of a Gaussian process, $l$ thus does not influence the value of MBR, but it influences the variance, i.e., the statistical error. We analyze VBR in the model system and for cell data, and find its minimum in dependence of $t$, $\tau$, and $l$. We find that as a rule of thumb, VBR is minimal if $l$ is comparable to the square root of the mean squared displacement (MSD) evaluated at time $\tau$. Further, we show that VBR of cell data is incompatible with a Gaussian process. Thus, VBR serves as a marker for non-Gaussianity.

The manuscript is organized as follows:
In Section \ref{sec:MBR-def}, we define MBR and introduce the conditioning time $\tau$, the observation time $t$, and the length parameter $l$, and we summarize the known properties of MBR.
In Section \ref{sec:MBR-Cells-Model}, we discuss the typical $t$- and $\tau$-dependence of MBR curves observed experimentally in cells and  qualitatively reproduce these curves using the mentioned model system.
In Section \ref{sec:Eeff}, we analyze the relation between MBR and effective energy, for cells, and for the model system.
In Section \ref{sec:MBR-cut}, we introduce and analyze the variance of back relaxation for Gaussian processes analytically.
In Section \ref{sec:VBR-Cells-Model}, we discuss VBR in both experimental data and model simulations.

\section{Mean Back Relaxation: Definition and previous work}
\label{sec:MBR-def}
Consider a stochastic degree of freedom $x(t)$ at time $t$. The mean back relaxation (MBR) is (minus) the average of the ratio of the displacement $x(t)-x(0)$ and the displacement $x(0) - x(-\tau)$ \cite{muenker_accessing_2024,knotz_mean_2024},
\begin{align}
    \begin{split}   
    \MBR(\tau,t,l)= \int \dif x_{-\tau} \dif x_{0} \dif x_t \left( - \frac{x_t - x_0}{x_0 - x_{-\tau}} \right) \times \\
    \times \vartheta_l( x_0 - x_{-\tau}) W_3(x_t,t;x_0,0;x_{-\tau} , -\tau),
    \end{split}\label{eq:MBR}
\end{align}
with the joint three point probability $W_3$. $W_3(x_{t_3},t_3;x_{t_2},t_2,x_{t_1},t_1)$ is the probability of finding the particle at the positions $x_{t_3},x_{t_2},x_{t_1}$ at the respective times $t_3>t_2>t_1$. The factor $\vartheta_l(x_0 - x_{-\tau}) \equiv \frac{\theta (\abs{x_0 - x_{-\tau}} - l)}{Z}$  with Heaviside function $\theta(x)$ and $Z= \mean{\theta \left(\left\vert x(0) - x(-\tau) \right\vert - l \right)}$ avoids division by $0$ in Eq.~\eqref{eq:MBR}. By construction, it is normalized to unity, i.e., $\mean{\vartheta_l(x(0) - x(-\tau))} = 1$, and it also introduces a length scale $l$, the smallest value of  $|x_{0} - x_{-\tau}|$ taken into account in Eq.~\eqref{eq:MBR}, i.e., a cut off length. A minus sign is included in Eq.~\eqref{eq:MBR} to render MBR positive in case of a {\it back} relaxation, i.e., if the displacements have opposite sign. MBR as defined in Eq.~\eqref{eq:MBR} thus depends on the two times $t$ and $\tau$ and on the length $l$. 

We repeat, for completeness, the main findings of previous work. In Refs.~\cite{muenker_accessing_2024,knotz_mean_2024}, we showed that, if (i) $\langle x \rangle$ is finite  and time independent, i.e., $x$ has a stationary distribution with finite mean, and (ii) the process is ergodic with finite relaxation times, one finds, with detailed balance,
\begin{align}
    \lim_{t \to \infty} \MBR(\tau,t,l) \overset{\text{d.b.}}{=} \frac{1}{2}.
\label{eq:12}
\end{align}
Notably, Eq.~\eqref{eq:12} holds for any positive values of $l$ and $\tau$ \cite{knotz_mean_2024}.  
Deviations from the long time value $\frac{1}{2}$, under the given conditions, thus  mark  broken detailed balance. This has, e.g., been illustrated for active Brownian particles trapped in a harmonic potential \cite{knotz_mean_2024}.

For a stationary Gaussian process, the following relation between MBR and  mean squared displacement (MSD) $\Delta x^2(t) = \mean{(x(t) - x(0))^2}$ has been found, \cite{knotz_mean_2024}
\begin{align}
    \MBR(\tau,t,l) = \frac{1}{2} \left( 1 - \frac{\Delta x^2(t+\tau) - \Delta x^2(t)}{\Delta x^2(\tau)} \right).
    \label{eq:MBR-MSD}
\end{align}
Remarkably, as seen on the rhs of Eq.~\eqref{eq:MBR-MSD}, for a Gaussian process,  MBR is independent of the length $l$.

For a probe particle immersed in a cell, as analyzed in Ref.~\onlinecite{muenker_accessing_2024} and below, a mean position $\mean{x}$ cannot be claimed to exist, so that no rigorous relation between a deviation of MBR from $\frac{1}{2}$ and broken detailed balance can be argued for in the remainder of this manuscript. In experiments in Ref.~\onlinecite{muenker_accessing_2024}, we have, however, both experimentally, as well as in a model system, observed a striking relation between effective energy of Eq.~\eqref{FDT:adapted} and the long time value of MBR. In Ref.~\onlinecite{muenker_accessing_2024}, we considered values of $\tau $ and $l $ both chosen small, i.e., close to  experimental temporal and spatial resolution. In this manuscript we extend the analysis to finite $\tau$. We also extend the mentioned model system to include memory, which yields a very good qualitative agreement with our experiments.

\section{Mean Back Relaxation: Model and experiment}
\label{sec:MBR-Cells-Model}
In this section, we will analyze the dependence of MBR on the parameters $\tau$ and $t$. We will initially show results from probe particles in living cells, and then introduce and analyze a model system which allows to rationalize the behavior observed for cells.
\begin{figure*}
    \centering    
    \includegraphics[width=\linewidth]{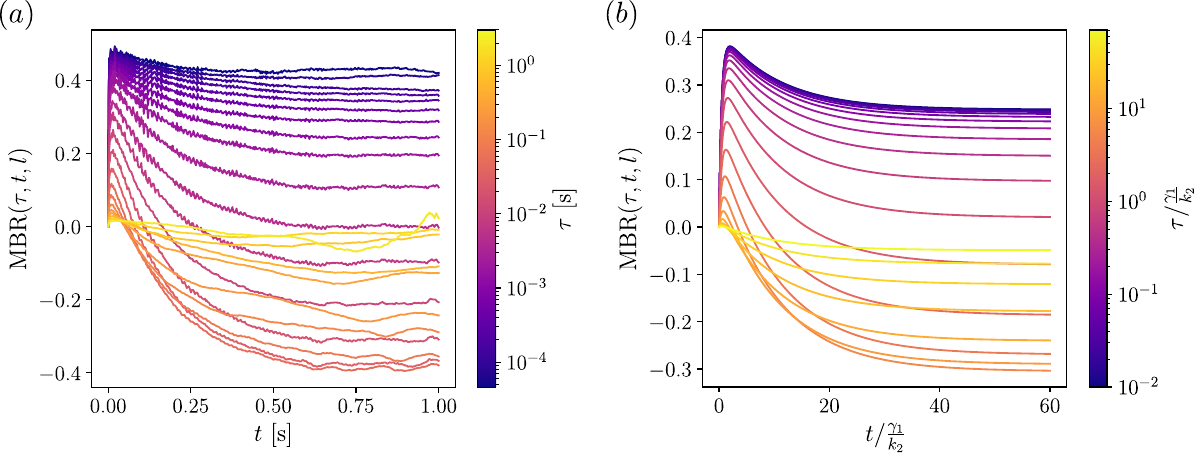}
    \caption{(a) MBR obtained from experimental data on A549 cells as a function of time $t$, for various values of $\tau$ and $l = \SI{0.002}{\micro\meter} $. $\tau$ ranges from \SI{0.05}{\milli\second} to \SI{3}{\second}. MBR reaches a plateau for $t\approx 0.5$s.  With increasing $\tau$, this plateau decreases until it reaches a minimum around $\tau \approx$ \SI{0.04}{\second} and then decays to zero for larger $\tau$. (b) MBR obtained from the RHC model of Eq.~\eqref{eq:model}, plotted in a similar manner as the data in a), for parameters $\frac{k_1}{k_2} = 1.0, \frac{\gamma_1}{\gamma_2} = 0.2, \frac{D_q}{D_1} = 0.5$, corresponding to  timescales of $\lambda_1^{-1} = 0.475 \frac{\gamma_1}{k_2}$ and $\lambda_2^{-1} = 10.525 \frac{\gamma_1}{k_2}$. The phenomenology is similar to what is observed in a), with the final approach of zero for large $\tau$ as a power law $\propto \frac{1}{\tau}$, see main text.  }
    \label{fig:MBR-Cells_HCM}
\end{figure*}

\subsection{Experiment: Cells}

\begin{figure*}
    \centering   
    \includegraphics[width=\linewidth]{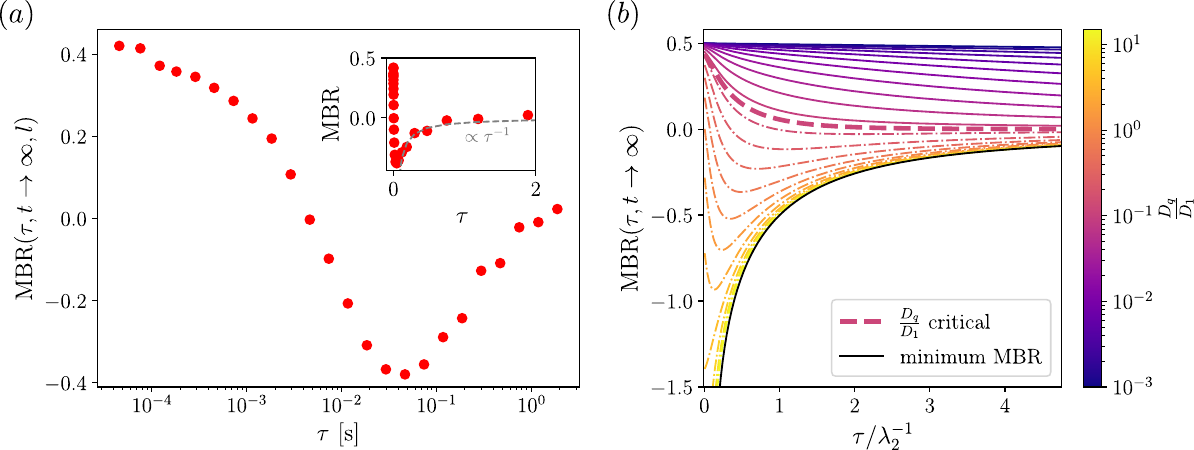}
    \caption{(a) Experiments: MBR$(t,\tau,l)$ at $t=$\SI{1}{\second} and $l=\SI{0.002}{\micro\meter}$ for A549 cells as a function of time $\tau$. For increasing $\tau$, MBR decreases, gets negative and decays to 0 from below. The minimum is approximately reached at $\tau \approx \SI{0.04}{\second}$. In the main panel, the x-axis is in log scale, and in the inset, we see the same data but in linear scale. (b) RHC Model: The long time limit, $t \to \infty$, of MBR as a function of $\tau$ for different activity $\frac{D_q}{D_1}$ as labeled.   $\frac{k_1}{k_2} = 1.0$ and $\frac{\gamma_1}{\gamma_2} = 0.2$ corresponding to the timescales of $\lambda_1^{-1} = 0.475 \frac{\gamma_1}{k_2}$ and $\lambda_2^{-1} = 10.525 \frac{\gamma_1}{k_2}$. Initially, MBR decreases with $\tau$. For large $\tau$, $\tau\gg\lambda_i^{-1}$, the curve   follows a power law $\frac{1}{\tau}$, see Eq.~\eqref{eq:HCM-MBR-LT-tauinfty}. For $\frac{D_q}{D_1}$ small, MBR remains positive, approaching zero from above. For large $\frac{D_q}{D_1}$, MBR approaches zero from below. The dashed line, showing Eq.~\eqref{eq:HCM-act-crit}, is the critical activity that separates the two cases. The black line shows the $\tau$ dependent minimum MBR, reached  for $\frac{D_q}{D_1} \to \infty$, and given in Eq.~\eqref{eq:MBR-Min}.}
    \label{fig:MBR-Cell_HCM-LT}
\end{figure*}

In Ref.~\onlinecite{muenker_accessing_2024}, we analyzed MBR from probes in various different living cell types, all showing similar characteristic behavior of MBR. Here we focus on one specific type of cell, namely A549 human lung carcinoma cells. Figure~\ref{fig:MBR-Cells_HCM}(a) shows MBR as a function of time $t$ for various different values of $\tau$, and $l = \SI{0.002}{\micro\meter} $. Recall that the curve shown in Ref.~\onlinecite{muenker_accessing_2024} corresponds to $\tau=0.6$ms. 
For all values of $\tau$ shown, MBR rises as a function of time, reaches a maximum, and then settles to $t$-independent value \cite{muenker_accessing_2024}. While qualitatively true for all $\tau$, the height of the maximum as well as the long time value are strongly dependent on $\tau$. Strikingly, the long time value is negative for intermediate values of $\tau$. For increasing $\tau$-values, MBR eventually vanishes for all times $t$, as the displacement in the period $[-\tau;0]$ has less and less influence on the displacement in the period $[0;t]$, so that the latter averages to zero in the accessible time frame.

Figure~\ref{fig:MBR-Cell_HCM-LT}(a) shows the long time value (here taken at $t=1$s) as a function of $\tau$, further highlighting the mentioned behavior, with a minimum at around $40$ms, and a gradual approach toward zero for larger $\tau$. To understand the behavior seen in Figs.~\ref{fig:MBR-Cells_HCM}(a) and \ref{fig:MBR-Cell_HCM-LT}(a), we introduce a model system, which will be shown to yield a similar MBR behavior. 

\subsection{Model system: "Random Horse and Cart"}
\label{sec:Diffusing potential with memory}
\subsubsection{Model}

\begin{figure}
    \centering
    \includegraphics[width=0.8\linewidth]{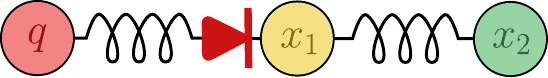}
    \caption{Sketch of the "Random Horse and Cart" (RHC) model of Eq.~\eqref{eq:model}.  The particle $x_1$ (the cart) is linearly coupled to the particle $q$ (the horse). This coupling is nonreciprocal, as $x_1$ feels the coupling force, but $q$ does not (the diode-spring). This non-reciprocity causes the system to be out of equilibrium. Further, $x_1$ is linearly coupled to a bath particle $x_2$ that mimics a viscoelastic environment. All particles follow overdamped Langevin equations.} 
    \label{fig:HCM-sketch}
\end{figure}

We continue by investigating the dependence of MBR on $t$ and $\tau$ in more depth in a Gaussian model system which can be driven away from equilibrium. Recall that for Gaussian systems, MBR does not depend on $l$. Thus, this model can not provide any insights into a possible dependence on $l$ in our experiments. We now extend the model introduced in Ref.~\onlinecite{muenker_accessing_2024} by an additional degree of freedom $x_2$ in order to introduce memory. The observed tracer particle at position $x_1$ (the "cart") has a bare friction coefficient $\gamma_1$, and it is trapped in a harmonic potential with stiffness $\kappa_2$, centered at position $q$. The potential may mimic an elastic coupling or caging of the particle in the cytoskeleton. The position $q$ performs a diffusion process with diffusion coefficient $D_q$, independent of the dynamics of $x_1$, thereby driving particle $x_1$ out of equilibrium (pulling the cart as a "horse"), thus mimicking the active dynamics of a cell environment. In addition to the model introduced in Ref.~\onlinecite{muenker_accessing_2024}, the tracer particle is here also linearly coupled, with coupling strength $k_1$, to a bath particle at position $x_2$ \cite{siegle_markovian_2010,muller_properties_2020, doerries_correlation_2021,khan_trajectories_2014,ginot_recoil_2022,caspers_how_2023,basu_universal_2024} with friction coefficient $\gamma_2$, which mimics the viscoelastic surrounding of the cytoplasm of a cell. The Langevin equations of this Random Horse and Cart (RHC) system are 
\begin{subequations}\label{eq:model}
\begin{align}
    \dot x_1 &= -\frac{k_1}{\gamma_1} (x_1 - x_2) - \frac{k_2}{\gamma_1} (x_1 - q) + \xi_1(t),\\
    \dot x_2 &= - \frac{k_1}{\gamma_2} (x_2 - x_1) + \xi_2(t),\\
    \dot q &= \xi_q(t),
\end{align}
\end{subequations}
with white noises $\mean{\xi_i(t) \xi_j(t^\prime)} = 2D_i \delta(t-t^\prime) \delta_{ij}$. The system is also shown as a sketch in Fig.~\ref{fig:HCM-sketch}.  $D_1=k_BT/\gamma_1$ and $D_2=k_BT/\gamma_2$ obey the Einstein relation with thermal energy $k_BT$. The trap diffuses with $D_q$. As the motion of the trap is not influenced by the position $x_1$,  the interaction between $q$ and $x_1$ is non-reciprocal. This model thus breaks detailed balance for finite $D_q$. We refer to $D_q$ as "activity" in the following. Notably, Eqs.~\eqref{eq:model} can also be obtained by starting with three particles connected via reciprocal interactions with a temperature and friction coefficient for particle $q$. Letting these two parameters go to infinity while keeping their ratios fixed yields  Eqs.~\eqref{eq:model}. The horse may thus be understood as a particle with high temperature and large friction coefficient.     
 \subsubsection{Mean squared displacement}
 \label{subsec:msd}
We first derive the mean squared displacement (MSD) of particle $x_1$ of the RHC, and then use Eq.~\eqref{eq:MBR-MSD} to find MBR.  MSD can be determined analytically and reads \cite{hoper_mean_2023} (see \ref{subsec:Methods:HCM})
\begin{align}
    \begin{split}   
    \beta k_2 \Delta x_1^2(t) = &2\frac{D_q}{D_1} \frac{k_2}{\gamma_1} t \\
    &+ \Psi(\lambda_1,\lambda_2) \left(1 - e^{-\lambda_1 t} \right)\\
    &    +\Psi(\lambda_1,\lambda_2) \left(1 - e^{-\lambda_2 t} \right),
    \end{split}
    \label{eq:HCM-MSD}
\end{align}
with rates 
\begin{align}
\begin{split}
    \lambda_{1,2} = \frac{k_2}{\gamma_1} \frac{1}{2} &\left[ 1 + \frac{k_1}{k_2}+ \frac{\gamma_1}{\gamma_2} \frac{k_1}{k_2} \right.\\
    &\left. \pm \sqrt{ \left(1 + \frac{k_1}{k_2} + \frac{\gamma_1}{\gamma_2} \frac{k_1}{k_2} \right)^2 - 4 \frac{\gamma_1}{\gamma_2} \frac{k_1}{k_2} } \right].
    \end{split}
\end{align}
W.l.o.g., we use $\lambda_1 > \lambda_2$ throughout. The rates are proportional to the bare rate $\frac{k_2}{\gamma_1}$ and dependent on dimensionless ratios otherwise. Notably, the rates are independent of activity $D_q$. $D_q$ enters MSD in the dimensionless factors  $\Psi(\lambda_1,\lambda_2) = f(\lambda_1,\lambda_2) + \frac{D_q}{D_1} g(\lambda_1,\lambda_2)$. The dimensionless functions $f(\lambda_1,\lambda_2)> 0$ and $g(\lambda_1,\lambda_2) < 0$ are given  in Appendix~\ref{subsec:HCM-MBR}, and are independent of $D_q$. 
The RHC model thus shows short time diffusion, i.e., $\lim_{t\ll \lambda_1^{-1},\lambda_2^{-1}} \Delta x_1^2(t)=2D_1t$, and long time diffusion  $\lim_{t\gg \lambda_1^{-1},\lambda_2^{-1}} \Delta x_1^2(t)=2D_q t$; The latter result is intuitive as the particle, for long times, follows the trap position $q$. 

The activity  enters MSD only in the dimensionless ratio  $\frac{D_q}{D_1}$, for which three regimes can be identified corresponding to the signs of the functions $\Psi$:  For 
\begin{align}
	\frac{D_q}{D_1} < \frac{\gamma_1}{k_2} \frac{(\lambda_1 + \lambda_2)\lambda_2}{\frac{k_2}{\gamma_1}+ \lambda_1} \label{eq:cond1}
\end{align}
$\Psi(\lambda_1,\lambda_2) > 0$ and $\Psi(\lambda_2,\lambda_1) > 0$, therefore both exponential terms in Eq.~\eqref{eq:HCM-MSD} contribute positively to the MSD, and MSD has a negative curvature for all times $t$. For 
\begin{align}
	\frac{\gamma_1}{k_2}\frac{(\lambda_1 + \lambda_2) \lambda_1}{ \frac{k_2}{\gamma_1} + \lambda_2}<\frac{D_q}{D_1} \label{eq:cond2}
\end{align}
$\Psi(\lambda_1,\lambda_2) < 0$ and $\Psi(\lambda_2,\lambda_1) < 0$, therefore both exponential terms contribute negatively to the MSD. MSD has a positive  curvature for all times $t$. In between, i.e., for 
\begin{align}
	\frac{\gamma_1}{k_2} \frac{(\lambda_1 + \lambda_2)\lambda_2}{\frac{k_2}{\gamma_1}+ \lambda_1} <  \frac{D_q}{D_1} < \frac{\gamma_1}{k_2}\frac{(\lambda_1 + \lambda_2) \lambda_1}{ \frac{k_2}{\gamma_1} + \lambda_2} \label{eq:cond3}
\end{align}
$\Psi$ corresponding to the larger time scale $\lambda_2^{-1}$ is negative while $\Psi$ corresponding to the smaller timescale $\lambda_1^{-1}$ is positive. Eq.~\eqref{eq:cond3} turns out to be a necessary, but not sufficient condition for MSD to change from negative curvature for short times to positive curvature for larger times. 

 At the boundaries, where either of the $\Psi$ change sign and go through zero,  the system displays only a single timescale.  Note that Eqs.~\eqref{eq:cond1}, \eqref{eq:cond2} and \eqref{eq:cond3}  are independent of the rate $\frac{k_2}{\gamma_1}$ because all contributions cancel with the prefactor in $\lambda_{1}$ and $\lambda_2$. Thus, they only depend on the dimensionless ratios $\frac{\gamma_1}{\gamma_2}$ and $\frac{k_1}{k_2}$.

We will discuss in subsection \ref{subsec:tauto0} and \ref{subsec:finitetau}, how these regimes translate to behavior of MBR. 

\subsubsection{Mean back relaxation: Dependence on $t$ for $\tau\to0$}
\label{subsec:tauto0}
\begin{figure}
    \centering
    \includegraphics[width=\linewidth]{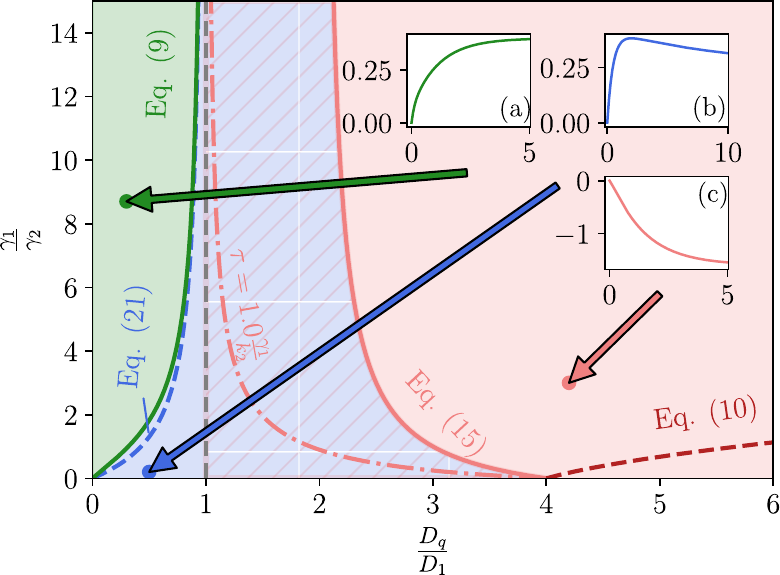}
	\caption{Regimes of qualitatively different shapes of MBR$(t,\tau)$ of the RHC model of Eq.~\eqref{eq:model}. Green regime:  MBR grows monotonically as a function of time $t$, see inset (a). In the red area, MBR shrinks monotonically as a function of time $t$ fir $\tau\to0$, see inset (c). In the blue area, MBR is non-monotonic as a function of time, for $\tau\to0$, see inset (b). The blue area with light red stripes corresponds to the regime, where, as a function of $\tau$, MBR turns from non-monotonic to monotonic in $t$, i.e., the red solid line moves to the left with increasing $\tau$. A red dashed dotted line indicates the boundary between non-monotonic and monotonic for a specific $\tau$ as labeled.    Long time value of MBR changes sign at $D_q/D_1=1$ shown as a gray dashed line.}
    \label{fig:HCM-MBR-typical}
\end{figure}

We discuss the behavior of MBR using Eq.~\eqref{eq:MBR-MSD}, and the solution for MSD of Eq.~\eqref{eq:HCM-MSD}.
We start by discussing the dependence of MBR on time $t$ in the limit of $\tau\to0$, which can be taken analytically for this model,
    \begin{align}
    \begin{split}    
     &\lim_{\tau\to0}\MBR(\tau,t) =\\
     &\frac{1}{2} \frac{\lambda_1\Psi(\lambda_1,\lambda_2) \left(1 - e^{-\lambda_1 t} \right) + \lambda_2\Psi(\lambda_2,\lambda_1) \left(1 - e^{-\lambda_2 t} \right)}{2 \frac{D_q}{D_1} \frac{k_2}{\gamma_1} +\lambda_1\Psi(\lambda_1,\lambda_2)  + \lambda_2\Psi(\lambda_2,\lambda_1) }.
     \end{split}
\end{align}
For the discussion of MBR, it is useful to remark, that for the given model, Eq.~\eqref{eq:MBR-MSD} turns in this limit to
\begin{align}
  \lim_{\tau\to0}  \MBR(\tau,t) = \frac{1}{2} \left( 1 - \frac{1}{2D_1}\frac{\partial}{\partial t}\Delta x^2(t)\right).
    \label{eq:MBR-MSD_tauto0}
\end{align}
MBR in this limit, as a function of time $t$, has three characteristic shapes, related to the cases discussed in Sec.~\ref{subsec:msd}. Fig.~\ref{fig:HCM-MBR-typical} displays these cases. Recall that the boundaries between the regimes depend only on $k_1/k_2$ and $\gamma_1/\gamma_2$, and Fig.~\ref{fig:HCM-MBR-typical} displays them for $k_1/k_2=1$ as a function of $\gamma_1/\gamma_2$.

For {\it small} $D_q/D_1$, i.e., for Eq.~\eqref{eq:cond1} fulfilled, MBR grows monotonically for all times $t$: This results from Eq.~\eqref{eq:MBR-MSD_tauto0} with the curvature of MSD being negative, i.e., $\frac{\partial}{\partial t}\Delta x^2(t)\leq 2D_1$ for all times $t$, and monotonically shrinking. 
For {\it large} $D_q/D_1$, i.e., for Eq.~\eqref{eq:cond2} fulfilled, MBR shrinks monotonically for all times $t$: This also results from Eq.~\eqref{eq:MBR-MSD_tauto0} with the curvature of MSD being positive, i.e., $\frac{\partial}{\partial t}\Delta x^2(t)\geq 2D_1$ for all times $t$, and monotonically growing. The two boundaries of Eqs.~\eqref{eq:cond1} and \eqref{eq:cond2} are shown in Fig.~\ref{fig:HCM-MBR-typical} as a dark green and dark red line, respectively.

In between these two lines, i.e., with Eq.~\eqref{eq:cond3} fulfilled, MBR curves can be nonmonotonic. Notably, Eq.~\eqref{eq:cond3} is a necessary condition for non-monotonic MBR to occur, but not a sufficient one. Searching for sufficient conditions, we find that a maximum  occurs for a subspace of Eq.~\eqref{eq:cond3}, namely for 
\begin{align}
    \frac{\gamma_1}{k_2}\frac{(\lambda_1 + \lambda_2)\lambda_2}{\frac{k_2}{\gamma_1}+ \lambda_1}  < \frac{D_q}{D_1} < -\frac{\lambda_1^2 f(\lambda_1,\lambda_2) + \lambda_2^2 f(\lambda_2,\lambda_1)}{\lambda_1^2 g(\lambda_1,\lambda_2) + \lambda_2^2 g(\lambda_2,\lambda_1)} .\label{eq:condMBR1}
\end{align}
Within these boundaries, MBR is positive for small $t$ and shows a maximum at a time of $ t_\text{max} = \frac{1}{\lambda_1 - \lambda_2} \log \left( - \frac{\lambda_1^2 \Psi(\lambda_1, \lambda_2)}{\lambda_2^2 \Psi(\lambda_2, \lambda_1)} \right)$. 
The lower bound of  Eq.~\eqref{eq:condMBR1} agrees with the one Eq.~\eqref{eq:cond1}, thus shown as a dark green line. The upper bound of Eq.~\eqref{eq:condMBR1} is shown as a light red curve in Fig.~\ref{fig:HCM-MBR-typical}. On the right hand side of this curve, i.e., for
\begin{align}
    -\frac{\lambda_1^2 f(\lambda_1,\lambda_2) + \lambda_2^2 f(\lambda_2,\lambda_1)}{\lambda_1^2 g(\lambda_1,\lambda_2) + \lambda_2^2 g(\lambda_2,\lambda_1)}  < \frac{D_q}{D_1}
    \label{eq:condMBR2}
\end{align}
fulfilled, MBR is negative and decreases monotonically, i.e., Eq.~\eqref{eq:cond2} is a sufficient but not necessary condition for such behavior. Notably, as is the case for the other boundaries, Eqs.~\eqref{eq:condMBR1} and \eqref{eq:condMBR2} only depend on the ratios $\frac{\gamma_1}{\gamma_2}$ and $\frac{k_1}{k_2}$.

Also, the long time value of MBR takes a simple form which is not influenced by the presence or absence of the particle $x_2$ in Eq.~\eqref{eq:model}, i.e., the long time value is identical to the one found for the case of $k_1=0$ studied in Ref.~\onlinecite{muenker_accessing_2024},
\begin{align}
    \MBR(\tau \to 0,t \to \infty) = \frac{1}{2} \left(1 - \frac{D_q}{D_1} \right).\label{eq:ttinf}
\end{align}
The order of limits is not important in Eq.~\eqref{eq:ttinf}, i.e., Eq.~\eqref{eq:ttinf} is found by either order. The long time value of MBR thus solely depends on the ratio $D_q/D_1$, changing sign at $D_q/D_1=1$, which is shown as a dashed line in Fig.~\ref{fig:HCM-MBR-typical}. 

In the following, we focus on parameters for which MBR shows a maximum as a function of $t$, as this is the behavior observed in the cell data in Fig.~\ref{fig:MBR-Cells_HCM}(a). See the top curve in Fig.~\ref{fig:MBR-Cells_HCM}(b), which corresponds to $\tau\to 0$.  Remarkably, for the parameters chosen, it looks very similar to the top curve in Fig.~\ref{fig:MBR-Cells_HCM}(a). As the simpler model of Ref.~\onlinecite{muenker_accessing_2024} shows no maximum as a function of time, we see that the particle $x_2$, i.e., memory, is crucial in obtaining qualitative agreements with cell data. 

\subsubsection{Dependence on $\tau$}
\label{subsec:finitetau}
We continue with analyzing the dependence of MBR on time $\tau$.  Eq.~\eqref{eq:condMBR2}  above, i.e., the boundary between non-monotonic and monotonic dependence on time $t$ can be extended for finite $\tau$. For finite $\tau$, curves are negative and monotonic in $t$, if
\begin{align}
\begin{split} 
    &- \frac{\lambda_1 f(\lambda_1,\lambda_2) \left(1 - e^{-\lambda_1 \tau} \right) + \lambda_2 f(\lambda_2,\lambda_1) \left(1 - e^{-\lambda_2 \tau} \right)}{\lambda_1 g(\lambda_1,\lambda_2) \left(1 - e^{-\lambda_1 \tau} \right) + \lambda_2 g(\lambda_2,\lambda_1) \left(1 - e^{-\lambda_2 \tau} \right)} \\
   & < \frac{D_q}{D_1}.
    \end{split}\label{eq:lhs}
\end{align}
Notably, $\tau$ enters this expression exponentially in the combinations $\tau \lambda_i$. With increasing $\tau$, the left hand side of Eq.~\eqref{eq:lhs} decreases, i.e., the curve separating blue and red areas in Fig.~\ref{fig:HCM-MBR-typical} moves to the left.  For $\tau \gg \frac{1}{\lambda_i}$, this line approaches   $\frac{D_q}{D_1} = 1$. 

The green line in Fig.~\ref{fig:HCM-MBR-typical}, given by  Eq.~\eqref{eq:cond1}, i.e., the "left" border of non-monotonic behavior, is independent of $\tau$. The area of nonmonotonic MBR behavior thus stays of finite size for any $\tau$. In Fig.~\ref{fig:HCM-MBR-typical}, we marked the area 
\begin{align}
 1<  \frac{D_q}{D_1} <-\frac{\lambda_1^2 f(\lambda_1,\lambda_2) + \lambda_2^2 f(\lambda_2,\lambda_1)}{\lambda_1^2 g(\lambda_1,\lambda_2) + \lambda_2^2 g(\lambda_2,\lambda_1)}   
\end{align}
by red stripes. In this area, MBR changes from non-monotonic for small $\tau$, to monotonic negative behavior for larger $\tau$. On the other hand, parameters in the blue area left of  $\frac{D_q}{D_1}=1$ are non-monotonic for all values of $\tau$.

Figure~\ref{fig:MBR-Cells_HCM}(b) shows curves for parameters in the latter case, i.e., in the blue area, as a function of $t$ for various values of $\tau$. For small $\tau$, $\tau\ll \lambda_i^{-1}$, the curve does not depend on $\tau$, and it is non-monotonic by construction, and it ends with a positive value for $t\to\infty$, as stated by Eq.~\eqref{eq:ttinf}.     

For $\tau$ comparable to $\lambda_1^{-1}$ and $\lambda_2^{-1}$, MBR decreases and the long time value eventually turns negative (Eq.~\eqref{eq:ttinf} holds only for $\tau\to0$). For large $\tau$, $\tau\gg \lambda_i^{-1}$, it asymptotically vanishes with a power law in $\tau$, which will be discussed in more detail below. The astonishing similarity with curves obtained from cells, shown in panel Fig.~\ref{fig:MBR-Cells_HCM}(a) suggests that the experimental system is located in the blue regime of Fig.~\ref{fig:HCM-MBR-typical}. 

\begin{figure}
    \centering    
    \includegraphics[width= \linewidth]{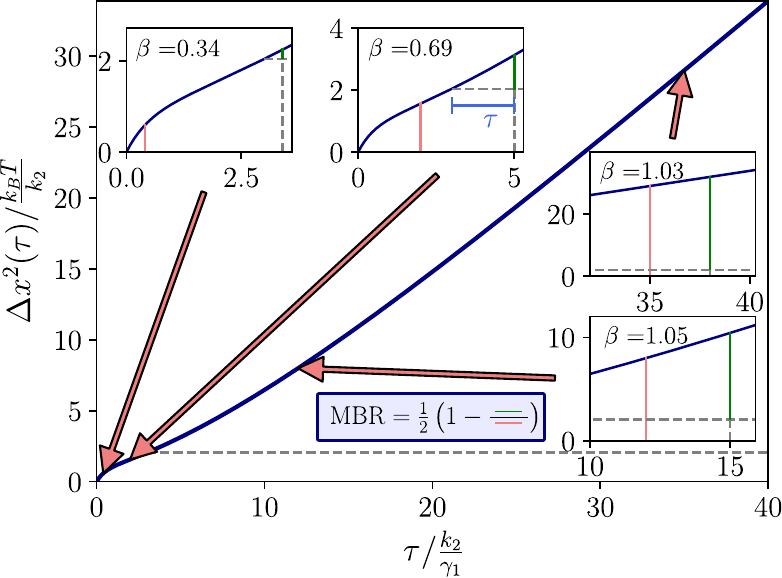}
    \caption{Illustration of MBR as resulting from MSD of RHC, at $t = 1 \frac{\gamma_1}{k_2}$, for different $\tau$. Arrows indicate the MSD at different $\tau$ values, and insets at the beginning of the arrows show construction of the ratio $\beta$ (with $\MBR(\tau,t) = \frac{1}{2} (1 - \beta)$, see main text) at the corresponding value of $\tau$. In the insets, the lengths of the red lines equal  $\Delta x^2(\tau)$ and the length of the green lines equal $\Delta x^2(t+\tau) - \Delta x^2(\tau)$. $\beta = \frac{\Delta x^2(t+\tau) - \Delta x^2(t)}{\Delta x^2(\tau)}$ is thus the ratio between the length of green and red lines. For small $\tau$ the red line is  longer then the green one, MBR is positive. For increasing $\tau$, the green line grows relatively larger compared to the red one, eventually leading to a negative MBR at intermediate $\tau$.  For  large $\tau$, the lengths of the two lines approach each other, and $\beta \to$ 1. MBR vanishes.}
    \label{fig:HCM-MBR} 
\end{figure}

The mentioned behavior is further visualized via Eq.~\eqref{eq:MBR-MSD} above, using the MSD of this model, in Fig.~\ref{fig:HCM-MBR}. Eq.~\eqref{eq:MBR-MSD} can be written $\MBR(\tau,t) = \frac{1}{2} (1 - \beta(\tau,t))$, with the ratio $\beta (\tau,t)= \frac{\Delta x^2(t+\tau) - \Delta x^2(t)}{\Delta x^2(\tau)}$. The numerator of $\beta$ is shown as a green line in the insets of Fig.~\ref{fig:HCM-MBR}, while the denominator is shown as a red line. For small $\tau$, MSD grows faster between $0$ and $\tau$ compared to the interval $t$ to $t+\tau$, hence the ratio $\beta$ is smaller then one: MBR is positive. With $\tau$ increasing $\beta$ increases as well, hence MBR gets smaller. When  $\beta $ exceeds unity, MBR is negative. For large $\tau$, long time diffusion dominates, the green and red lines approach each other in length, i.e., $\beta \to 1$ and MBR$\to0$. MBR thus vanishes for large $\tau$.

\subsubsection{Limit of $t\to\infty$ as a function of $\tau$}
The dependence on $\tau$ can be investigated easier in the limit of $t\to\infty$, in which case we find, 

\begin{align}
    &\MBR(\tau,t \to \infty) =  \notag\\ 
    &\frac{1}{2} \left(1 + \frac{ 2 \frac{D_q}{D_1} \frac{k_2}{\gamma_1} \tau}{\Psi(\lambda_1,\lambda_2) \left( 1 - e^{-\lambda_1 \tau} \right) + \Psi(\lambda_2,\lambda_1) \left( 1 - e^{-\lambda_2 \tau} \right)}\right)^{-1}
    \label{eq:HCM-MBR-LT-tau}
\end{align}

For $\tau\to0$, this expression approaches Eq.~\eqref{eq:ttinf}. Notably, in this expression, $\tau$ appears in the product $\tau\lambda_i$, but also in a power law form. The latter dominates in the limit of  $\tau\gg \lambda_i^{-1}$, 
\begin{align}
    \MBR(\tau\to\infty,t \to \infty) = \frac{1}{4} \frac{\Psi(\lambda_1,\lambda_2) + \Psi(\lambda_2,\lambda_1)}{\frac{D_q}{D_q} \frac{k_2}{\gamma_1} \tau}.
    \label{eq:HCM-MBR-LT-tauinfty}
\end{align}
For large $\tau$, the MBR vanishes as a power law, and not, as might be naively expected, exponentially.  

Figure~\ref{fig:MBR-Cell_HCM-LT}(b) shows the limit of $t\to\infty$  as a function of $\tau$ for various values of $D_q/D_1$.  The curves start at $\tau=0$, with the value given by Eq.~\eqref{eq:ttinf}. For large $\tau$, MBR approaches zero with the power law of  Eq.~\eqref{eq:HCM-MBR-LT-tauinfty}. The power law changes sign when $\Psi(\lambda_1,\lambda_2) + \Psi(\lambda_2,\lambda_1) = 0$, which corresponds to 
\begin{align}
    \frac{D_q}{D_1} = -\frac{f(\lambda_1,\lambda_2)  + f(\lambda_2,\lambda_1) }{g(\lambda_1,\lambda_2)  + g(\lambda_2,\lambda_1) }
    \label{eq:HCM-act-crit}
    \end{align}
Notably, at this value of  $D_q/D_1$, the power law vanishes, and MBR approaches zero exponentially with $\tau$. This critical activity as a function of $\frac{\gamma_1}{\gamma_2}$ is also shown as the blue dotted line in Fig.~\ref{fig:HCM-MBR-typical}.

Eq.~\eqref{eq:HCM-act-crit} and Eq.~\eqref{eq:ttinf} thus give rise to a range of activity values for which the large $t$ limit, as a function of time $\tau$, changes sign form positive to negative. This occurs for 

\begin{align}
    -\frac{f(\lambda_1,\lambda_2)  + f(\lambda_2,\lambda_1) }{g(\lambda_1,\lambda_2)  + g(\lambda_2,\lambda_1) } < \frac{D_q}{D_1} < 1\label{eq:signchange}
\end{align}

For very large activity, $D_q/D_1\to\infty$,  MBR$(\tau,t\to\infty)$ approaches a limiting line, given by 

\begin{align}
 &   \MBR_\text{min}(\tau,t\to\infty) \notag\\&= \frac{1}{2} \frac{g(\lambda_1,\lambda_2) (1 - e^{-\lambda_1 \tau}) + g(\lambda_2,\lambda_1) (1 - e^{-\lambda_2 \tau})}{2 \tau + g(\lambda_1,\lambda_2) (1 - e^{-\lambda_1 \tau}) + g(\lambda_2,\lambda_1) (1 - e^{-\lambda_2 \tau})}.
    \label{eq:MBR-Min}
\end{align}

In the RHC model of Eq.~\eqref{eq:model}, MBR cannot fall below this line.
This line is shown in Fig.~\ref{fig:MBR-Cell_HCM-LT} as a thick black line. For $\tau\to\infty$, it approaches zero as $\frac{g(\lambda_1,\lambda_2) + g(\lambda_2,\lambda_1)}{2 \tau}$. For   $\tau\to 0$, it diverges as $\frac{g(\lambda_1,\lambda_2) \lambda_1^2 + g(\lambda_2,\lambda_1) \lambda_2^2}{2 \tau}$, in agreement with the finding of Eq.~\eqref{eq:ttinf} that, in the limit $\tau\to0$, MBR can reach arbitrary negative values. 

The surprising agreement of the curves shown in Fig.~\ref{fig:MBR-Cell_HCM-LT}(a) and (b) is a further support that the simple model captures the system in a cell in many aspects qualitatively. Notably, the experimental curve in Fig.~\ref{fig:MBR-Cell_HCM-LT}(a) crosses from positive to negative, i.e., indicating to be in the range of Eq.~\eqref{eq:signchange}.

\section{MBR and effective energy}
\label{sec:Eeff}
As discussed in the introduction, we observed in Ref.~\onlinecite{muenker_accessing_2024} a surprising phenomenological relation between MBR and effective energy (see  Eq.~\eqref{FDT:adapted2} below for its definition). While in Ref.~\onlinecite{muenker_accessing_2024}, we restricted to small values of $\tau$, we will here  consider a range of $\tau$ values.

\subsection{Effective Energy and FDT}
The effective energy $E_\text{eff}(\omega)$ is defined as the ratio of correlation and response function at frequency $\omega$,
    \begin{align}
    E_\text{eff}(\omega) = \frac{\omega C(\omega)}{2\chi^{\prime \prime}(\omega)}.
    \label{FDT:adapted2}
\end{align}
For systems obeying detailed balance, FDT is recovered with $E_\text{eff} = k_B T$. Studying $E_\text{eff} $ is thus one way of quantifying the violation of FDT and of detailed balance. 


\subsection{Effective energy and MBR in RHC model system}
For the RHC model of Eq.~\eqref{eq:model}, response function and correlation in Eq.~\eqref{FDT:adapted} can be found analytically, see Appendix \ref{subsec:Methods:HCM} for the  detailed calculation \cite{zwanzig_nonequilibrium_2001,ayaz_generalized_2022,dalton_fast_2023}. Notably, the linear response function $\chi$ is independent of the activity parameter $D_q$, so that the dependence of the ratio of $\chi$ and $C$ on $D_q$ results purely from the correlation $C$. $E_\text{Eff}$ can thus be found as the ratio of correlation taken at $D_q$ and at $D_q=0$, 
\begin{align}
    \frac{E_\text{Eff}}{k_BT} &= \frac{\mean{x(\omega) x(-\omega)}}{\mean{x(\omega) x(-\omega)}_{D_q = 0}}\\
    &=1 + \frac{D_q}{D_1} \frac{k_2^2}{\gamma_1^2} \frac{1}{\omega^2} \frac{\frac{k_1^2}{\gamma_2^2} + \omega^2}{\frac{k_1^2}{\gamma_2^2} + \omega^2 + \frac{k_1^2}{\gamma_1 \gamma_2}}\\
     &\overset{\omega \to 0}{=}  1 + \frac{D_q}{D_1} \frac{k_2^2}{\gamma_1^2} \frac{1}{\omega^2} \frac{1}{1 + \frac{\gamma_2}{\gamma_1}}\\
      &=1 + \frac{E_0}{k_BT} \frac{\omega_c^2}{\omega^2}.
\end{align}
In the third line, we took the limit $\omega \to 0$ to identify the asymptotic behavior, and we introduced the "corner" frequency $\omega_c^2 = \frac{k_2^2}{\gamma_1^2}$. We extract the  amplitude 
\begin{align}
E_0 = k_B T \frac{D_q}{D_1} \frac{1}{1 + \frac{\gamma_2}{\gamma_1}}.\label{eq:E0}
\end{align}
$E_0$ depends on activity $\frac{D_q}{D_1}$ and on the ratio $\frac{\gamma_2}{\gamma_1}$ of the friction of bath and tracer particles. For $\gamma_2\ll\gamma_1$, we recover the case studied in Ref.~\onlinecite{muenker_accessing_2024}, where particle $x_2$ is absent, and $E_0 = k_B T D_q/D_1$.

\begin{figure*}
    \centering    
    \includegraphics[width=\linewidth]{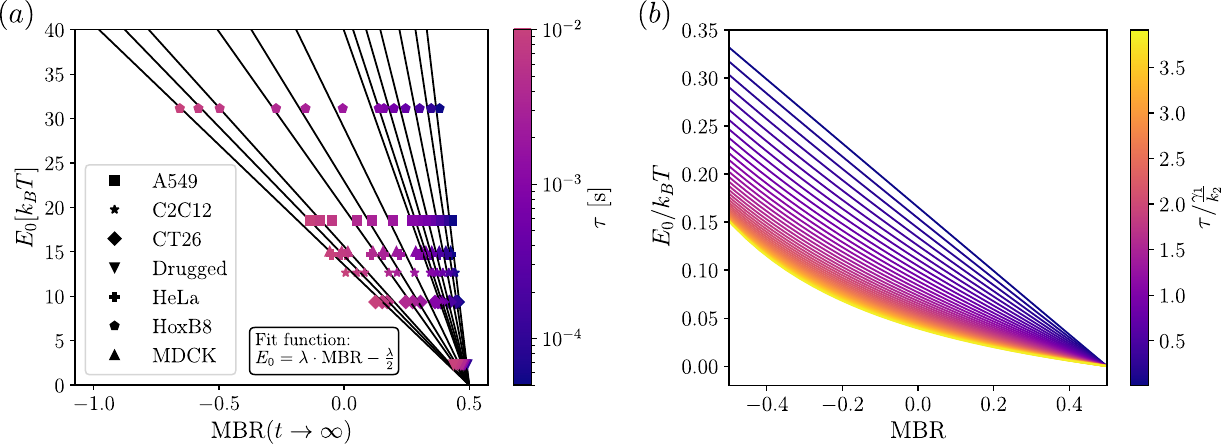}
    \caption{(a) Amplitude $E_0$ of effective energy against the long time limit of MBR for multiple $\tau$ between \SI{5e-5}{\second} and \SI{0.01}{\second}, for different cell types as labeled. For each $\tau$ value the values from different cells fall on a straight line, which is fitted and also shown. For increasing $\tau$ the slope of that line decreases. (b)  Effective energy amplitude $E_0$ against the long time limit of  MBR for $\frac{\gamma_1}{\gamma_2} = 0.2$ and $\frac{k_1}{k_2} = 1.0$ in the RHC model of Eq.~\eqref{eq:model}. Color indicates the respective $\tau$. For small $\tau$ the relation between $E_0$ and MBR is linear, and it is curved for larger $\tau$. Close to MBR$=1/2$, it is linear for any $\tau$, see main text.}
    \label{fig:E0_Tau}
\end{figure*}

Eq.~\eqref{eq:HCM-MBR-LT-tau} can be solved for 
 $D_q/D_1$ in terms of  MBR (we use for brevity, in the following equations, notation $M\equiv \MBR(\tau, t \to \infty)$),
\begin{align}
    \frac{D_q}{D} =  \frac{(1 -2M)(F_{12}  + F_{21})}{4 M \frac{k_2}{\gamma_1} \tau - (1-2M)(G_{12}  + G_{21} )}
    \label{eq:Dr-MBR0}
\end{align}
with abbreviations
\begin{align}
    F_{ij}&=f(\lambda_i,\lambda_j) (1 - e^{-\lambda_i \tau}),\\
    G_{ij}&=g(\lambda_i,\lambda_j) (1 - e^{-\lambda_i \tau}).
\end{align}
With Eq.~\eqref{eq:E0}, this yields a relation between $E_0$ and $M$,
\begin{align}
    \frac{E_0}{k_BT}=\frac{1}{1+\frac{\gamma_2}{\gamma_1}}  \frac{(1 -2M)(F_{12}  + F_{21})}{4 M \frac{k_2}{\gamma_1} \tau - (1-2M)(G_{12}  + G_{21} )}.
    \label{eq:Dr-MBR}
\end{align}
Figure~\ref{fig:E0_Tau}(b) shows $E_0$ as a function of $M$ from Eq.~\eqref{eq:Dr-MBR} for various values of $\tau$. In the limit $\tau \to 0$, Eq.~\eqref{eq:Dr-MBR} simplifies to the linear relation $\frac{E_0}{k_BT}=\frac{1}{1+\frac{\gamma_2}{\gamma_1}}  (1 -2M)$, which was displayed in Ref.~\onlinecite{muenker_accessing_2024} (for $\gamma_2\ll\gamma_1$). Such relation  is also observed for cell data. For finite $\tau$, the relation is curved, as seen in Fig.~\ref{fig:E0_Tau}(b). Even in these cases, for moderate deviations of MBR from $\frac{1}{2}$, a linear relation is found \begin{align}
\begin{split}
\frac{E_0}{k_BT} &= \frac{1}{1+\frac{\gamma_2}{\gamma_1}} \times\\ &\times \frac{e^{-\lambda_1\tau}(\lambda_2- \frac{k_2}{\gamma_1})-\lambda_2-(e^{-\lambda_2\tau}(\lambda_1-\frac{k_2}{\gamma_1})-\lambda_1)}{ \frac{k_2}{\gamma_1}(\lambda_1-\lambda_2)\tau}\times \\
    &\times(1-2M) +\mathcal{O}((1-2M)^2)
    \end{split}\\
    &= \frac{\lambda(\tau)}{k_B T} \left(M - \frac{1}{2} \right) + \mathcal{O}\left(\left( M -\frac{1}{2} \right)^2 \right).
    \label{eq:MBR-E0-slope-model}
\end{align}
The slope of this linear relation, $\lambda(\tau)$,  decreases with $\tau$, as is visible in Fig.~\ref{fig:E0_Tau}(b). 

\begin{figure*}
    \centering    
    \includegraphics[width=\linewidth]{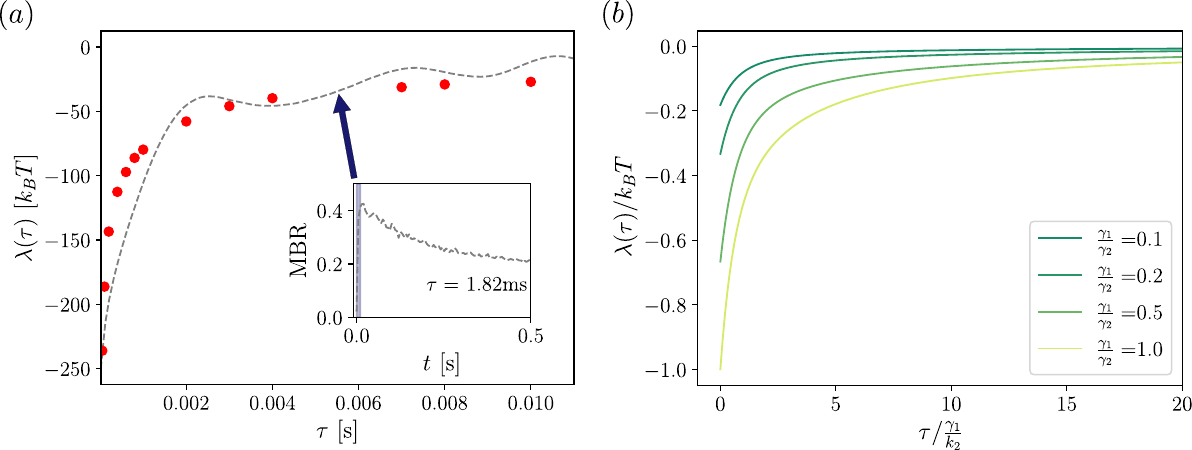}
    \caption{(a) Slope $\lambda(t)$ of the linear relation between $E_0$ and long time limit of MBR, extracted from Fig.~\ref{fig:E0_Tau}(a), against $\tau$. The magnitude  of the slope decreases with increasing $\tau$. The grey dotted line shows MBR as a function of time $t$ of Fig.~\ref{fig:MBR-Cells_HCM}a) for $\tau =  \SI{1.82}{\milli\second}$, shifted and rescaled along the $y$-axis, i.e., MBR in the blue area in the inset, which shows original MBR of Fig.~\ref{fig:MBR-Cells_HCM}a).   
    (b) Slope $\lambda(\tau)$ of the linear expansion in Eq.~\eqref{eq:MBR-E0-slope-model} for $\frac{k_1}{k_2} = 1$ and various $\frac{\gamma_1}{\gamma_2}$ as labeled. For large $\tau$, $\lambda$ decays as a power law, $\lambda\simeq -2\tau^{-1}/(1 + \gamma_2/\gamma_1)$ and for $\tau \to 0$ it approaches the constant $-2/(1 + \gamma_2/\gamma_1)$. }
    \label{fig:E0_Slope}
\end{figure*}

Figure~\ref{fig:E0_Slope}(b) shows the slope $\lambda(\tau)$ as a function of $\tau$, for various ratios of $\gamma_1/\gamma_2$, showing the mentioned decrease. For large $\tau$, $\tau \gg \{\lambda_1^{-1},\lambda_2^{-1}\}$, the slope decreases with a power law $\tau^{-1}$, i.e., in this limit,
\begin{align}
\begin{split}
\frac{E_0}{k_BT} &= -\frac{2}{1+\frac{\gamma_2}{\gamma_1}}  \frac{ \gamma_1}{k_2 \tau}  \left(M - \frac{1}{2} \right) + \mathcal{O}\left(\left( M -\frac{1}{2} \right)^2 \right).\label{eq:pl}
    \end{split}
\end{align} 
Interestingly, the prefactor of the power law in Eq.~\eqref{eq:pl} is very similar to the result of  $\lambda$ for $\tau\to0$ found above. 
What is the range of validity of the linear relation in Eq.~\eqref{eq:MBR-E0-slope-model}? It must break if  $\MBR$ approaches its minimum as a function of activity $\MBR_\text{min}$ given in Eq.~\eqref{eq:MBR-Min}.

In this case, increasing activity $\frac{D_q}{D}$ does no longer decrease $M$, as seen in Fig.~\ref{fig:MBR-Cell_HCM-LT}(b). As $E_0$ in Eq.~\eqref{eq:E0} increases with $D_q$, independent of $\tau$, the linear relation can than no longer hold. This lower bound for MBR  depends on $\tau$ and diverges in the limit $\tau\to 0$. A smaller $\tau$, therefore, increases the range where the long time value of MBR is sensitive to an increase of the driving $\frac{D_q}{D}$. 

For larger $\tau$, $M$ comes closer to its minimum, and the relation between $M$ and $E_0$ must become less precise. We will thus, in our experiments, restrict to the range where $M$ decreases as a function of $\tau$ (see Fig.~\ref{fig:MBR-Cell_HCM-LT}(a)).

\label{sec:Efftheo}
\subsection{Experiment: Effective Energy and MBR relation for various $\tau$}
\label{sec:effective-energy}
In a previous study \cite{muenker_accessing_2024}, we determined the effective energy and the viscoelastic properties of 8 different cell types and cell conditions using optical tweezers based active and passive microrheology (Material and Methods \ref{Methods: Active and passive microrheology}). In particular, we investigated HeLa cells, passivated HeLa cells, A549 cells, C2C12 cells, CT26 cells, HoxB8 cells, and MDCK cells (Material and Methods \ref{Methods: Cells}). This study yielded, for all cells investigated, good agreement with the phenomenological relation,
 \begin{align}
     E_\text{Eff} = E_0 \left(\frac{\omega_0}{\omega} \right) ^{\nu}+ k_B T,
 \end{align}
with $\nu \approx 1$ for all cell types. We chose  $\omega_0 =$\SI{1}{\hertz}, which defines and yields, for each cell type, a value for amplitude $E_0$. In Ref.~\onlinecite{muenker_accessing_2024}, we chose a small $\tau$-value of \SI{0.6}{\milli\second}  for evaluation of MBR, and observed a striking linear dependence between effective energy amplitude $E_0$ and MBR long-time value. Here we  investigate this relation for different $\tau$ values, limiting ourselves to $\tau\leq$\SI{0.01}{\second}; In Fig.~\ref{fig:MBR-Cell_HCM-LT}(a), we see that, in this regime, MBR decreases with $\tau$, see also the discussion at the end of Sec.~\ref{sec:Efftheo}.
Strikingly, when we compare the effective energy amplitude $E_0$ to MBR longtime value for different values of $\tau$, we  observe linear dependencies for all values of $\tau$, as can be seen in Fig.~\ref{fig:E0_Tau}(a). 

For each $\tau$-value, the relation between $E_0$ and MBR longtime value can be accurately fit  by a linear function (see Supplementary table \ref{si: r2 values e0 mbr fit} for fitting  uncertainties)
\begin{align}
    \label{fitunfunction mbr e0}
    E_0=\lambda(\tau) \left(\text{MBR}(t\to\infty,\tau)-\frac{1}{2}\right).
\end{align}
This result suggests that knowledge of the $\tau$-specific slope $\lambda(\tau)$, as shown in Fig. \ref{fig:E0_Slope}(a) is sufficient to determine the effective energy amplitude $E_0$ from an MBR measurement in these cases. As Fig. \ref{fig:E0_Slope}(a) shows, the magnitude of the slope decreases with increasing $\tau$, similarly as in the model system discussed in Sec.~\ref{sec:Efftheo} and shown in Fig.~\ref{fig:E0_Slope}(b). Figure \ref{fig:E0_Slope}(a) also shows, as a comparison, the time $t$ dependent $\MBR(t)$ of  Fig.~\ref{fig:MBR-Cells_HCM} a) (grey dotted line) shifted and rescaled along the $y$-axis. This shows that the time dependence of $\lambda(\tau)$ for small $\tau$ is comparable to the time  dependence of $\MBR(t)$ for small $t$. This allows to estimate, from MBR, on what time scales $\lambda(\tau)$ is expected to vary. Generally MBR can therefore also be used to scan relevant time scales in the system. Returning to the initial questions, we thus extend the relation between MBR and effective energy to finite values of $\tau$,  compared to Ref.~\onlinecite{muenker_accessing_2024}.


\section{Variance of Back Relaxation} 
\label{sec:MBR-cut}
For practical purposes it is important to estimate the statistical error of MBR. 
In this section, we address this by defining the   variance of back relaxation (VBR). We will also find the relation between VBR and MSD for a Gaussian process. We will then evaluate VBR in the model of Eq.~\eqref{eq:model} and for experimental data from cells. 
\subsection{Formal expression}
VBR is given by the variance of the observable averaged in Eq.~\eqref{eq:MBR},
\begin{align}
\begin{split}
    \VBR(\tau,t,l) &= \mean{\left( -\frac{x(t) - x(0)}{x(0) - x(-\tau)} \vartheta_l(x(0) - x(-\tau)) \right)^2}\\
    &- \mean{ \left(-\frac{x(t) - x(0)}{x(0) - x(-\tau)} \vartheta_l(x(0) - x(-\tau)) \right)}^2.
    \end{split}
\end{align}
\subsection{Gaussian process: Dependence on $t$, $\tau$ and $l$}

\begin{figure}
    \centering
    \includegraphics[width=\linewidth]{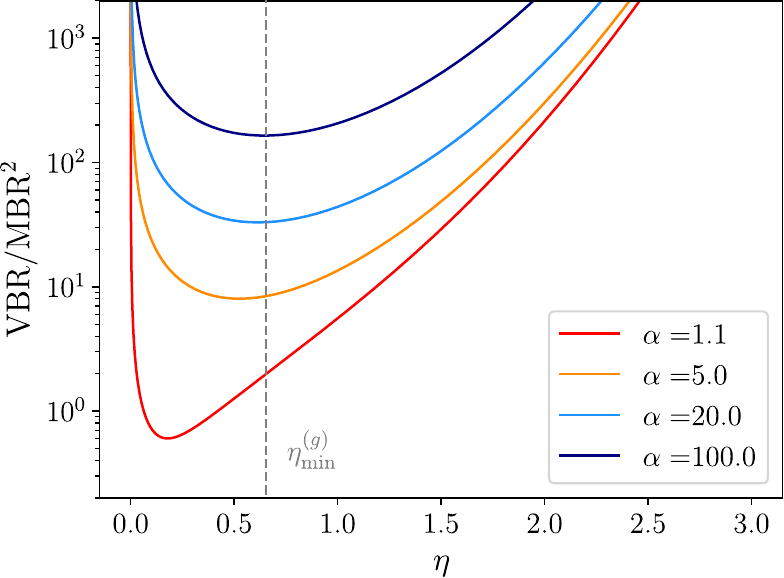}
    \caption{Variance of back relaxation (VBR) versus
    the length parameter $\eta = \frac{l}{\sqrt{2 \Delta x^2(\tau)}}$, from Eq.~\eqref{eq:Var_MBR_Gauss}, for various values of $\alpha = 4\left(1 - \frac{\Delta x^2(t+\tau) - \Delta x^2(t)}{\Delta x^2(\tau)} \right)^{-2} \frac{\Delta x^2(t)}{\Delta x^2(\tau)}$.
 Gray vertical line gives $\eta^{(g)}_\text{min}$, the position of the minimum of VBR approached for large $\alpha$.}
    \label{fig:cutoff}
\end{figure}

For a Gaussian process, VBR is  found in terms of MSD and MBR (see methods  \ref{Methods:Var-MBR} for details), 
    \begin{align}
    \VBR(\tau,t,l) 
    = \MBR(\tau,t)^2 \left[ (\alpha(\tau,t) - 1) g(\eta) +  h(\eta) \right].
    \label{eq:Var_MBR_Gauss}
\end{align}

We introduced the functions 
\begin{align}
g(\eta)&= \frac{1}{\text{erfc}(\eta)^2} \left( \frac{1}{\sqrt{\pi}} \frac{e^{-\eta^2}}{\eta} - \text{erfc}(\eta) \right),\\ h(\eta)&=\left( \frac{1}{\text{erfc}(\eta)} - 1 \right).\label{eq:h}
\end{align}
with the complementary error function $\text{erfc}(x)$. We also introduced  a dimensionless number 
\begin{align}
\alpha(\tau,t) =  \frac{1}{\MBR(\tau,t)^2}\frac{\Delta x^2(t)}{\Delta x^2(\tau)}.\label{eq:a} 
\end{align}
Notably,  for Gaussian systems $\alpha\geq 1$. 

We introduced the dimensionless length parameter $\eta = \frac{l}{\sqrt{2 \Delta x^2(\tau)}}$. Notably, for the Gaussian process, MBR is independent of $l$, but VBR in Eq.~\eqref{eq:Var_MBR_Gauss} depends on it.

 Note that MBR in Eq.~\eqref{eq:Var_MBR_Gauss} can be replaced in terms of MSD via Eq.~\eqref{eq:MBR-MSD}, so that VBR may equivalently be expressed purely in terms of MSD.

For any $\tau$ and $t$, VBR in Eq.~\eqref{eq:Var_MBR_Gauss} has a minimum as a function of $\eta$: $g(\eta\to\infty)\to  \frac{\sqrt{\pi}}{2} \frac{e^{\eta^2}}{\eta}$, and $h(\eta\to\infty)\to \sqrt{\pi} \eta e^{\eta^2}$, i.e., VBR diverges as $\eta \to\infty$ as long as MBR is nonzero. For $\eta \to 0$, $g$ diverges as $ g(\eta\to0)\to\frac{1}{\sqrt{\pi}} \frac{1}{\eta} $ while $h(\eta\to0)\to 0$ stays finite. VBR thus diverges as a power law for $\eta\to0$, except for special cases $\text{MBR}=0$ or $\alpha=1$. 

The divergence of VBR for large $\eta$ is understood from the observation that  $\mean{\theta (\abs{x_0 - x_{-\tau}} - l)}$ goes to zero in that limit. In other words, for $\eta\to \infty$, there are less and less occurrences of displacements $|x_0-x_{-\tau}|$ larger than $l$, yielding worse and worse statistics. For small $\eta$, events with smaller and smaller $|x_0-x_{-\tau}|$ are taken into account in Eq.~\eqref{eq:MBR}, yielding a larger variance due to the smaller and smaller denominator in that equation.     

The minimum of VBR as a function of $t$, $\tau$ and $l$ is of interest, as one may expect the smallest statistical error of MBR at that minimum.  We will continue by discussing this minimum as a function of $\eta$, starting with $t$ and $\tau$ fixed. Notably, $g(\eta)$ shows a global minimum at ${\eta}_\text{min}^{(g)} = 0.654334 $, corresponding to a length $l$ of ${l}\approx 0.92537 \sqrt{\Delta x^2(\tau)}$, i.e., $l$ comparable to the square root of MSD at time $\tau$. In contrast, $h(\eta)$ in Eq.~\eqref{eq:h} is a monotonically growing function of $\eta$, and the minimum of VBR, $\eta_{\rm min}$, thus depends on $\alpha$ in Eq.~\eqref{eq:Var_MBR_Gauss}, and is smaller than  ${\eta}_\text{min}^{(g)}$, $\eta_{\rm min}\leq {\eta}_\text{min}^{(g)}$. In practical situations, $\text{MBR}\sim \mathcal{O}(1)$ and if $\Delta x^2(t)\gg \Delta x^2(\tau)$,  $\alpha-1\gg 1$. We expand around this case, i.e., perform an asymptotic expansion around $\alpha-1$ infinite, finding, for the minimum of VBR, 

\begin{align}
  \eta_\text{min}(\alpha) = \eta_\text{min}^{(g)} - \frac{h^\prime \left(\eta_\text{min}^{(g)} \right)}{g^{\prime  \prime} \left(\eta_\text{min}^{(g)} \right)} \frac{1}{\alpha - 1} + \mathcal{O} \left( \left( \alpha - 1 \right)^{-2} \right), 
    \label{eq:Approx-min-Var}
\end{align}
with expansion coefficients $h^\prime \left({\eta}_\text{min}^{(g)} \right) = 5.84259$ and $g^{\prime \prime} \left({\eta}_\text{min}^{(g)} \right) = 6.71207$. For $\alpha\to\infty$, $\eta_{\rm min}$ approaches $\eta_\text{min}^{(g)}$, and is smaller for finite $\alpha$.  

Figure~\ref{fig:cutoff} shows VBR as a function of $\eta$ for several values of $\alpha$. For all curves shown, VBR has a single minimum. Both the value of $\eta_{\rm min}$ as well as the minimal value of VBR decrease with $\alpha$. For large $\alpha$, the position $\eta_{\rm min}$ is seen to become independent of $\alpha$ (Eq.~\eqref{eq:Approx-min-Var}), while VBR, at that minimum, grows linearly in $\alpha$, as expected from Eq.~\eqref{eq:Var_MBR_Gauss}.

\begin{figure}
    \centering
    \includegraphics[width=\linewidth]{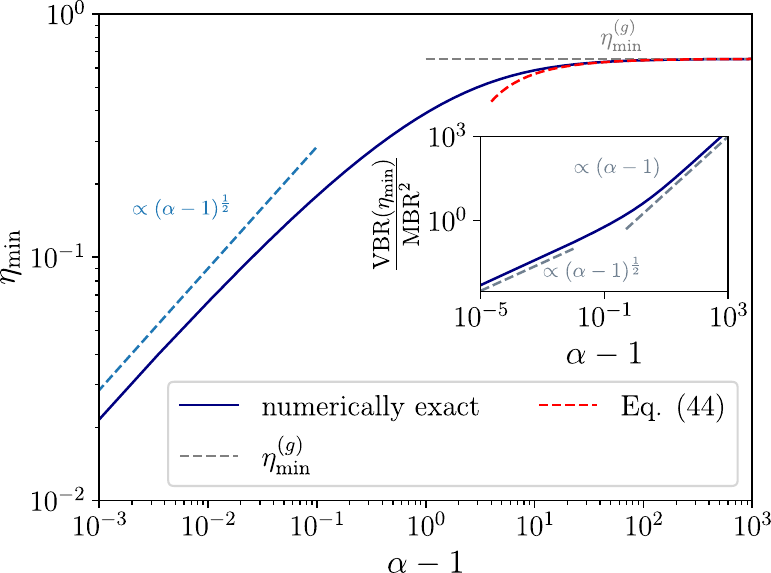}
    \caption{Position of minimum of VBR, $\eta_{\rm min}$, as a function of $\alpha$. Solid blue line shows the numerically exact solution, and the dotted grey and red lines show the zeroth and first order terms of Eq.~\eqref{eq:Approx-min-Var}, respectively. Inset shows VBR, evaluated at $\eta_{\rm min}$, as a function of $\alpha-1$, displaying the two power laws with a cross over at $\alpha - 1\approx 0.1$.}
 
    \label{fig:Var-cut-num-1st}
\end{figure}

As is evident in Eq.~\eqref{eq:Var_MBR_Gauss},  $\eta_{\rm min}$ depends only on $\alpha$, and 
Fig.~\ref{fig:Var-cut-num-1st} shows that dependence. For large $\alpha$, $\eta_\text{min}$ approaches  ${\eta}_\text{min}^{(g)}$ with a power law, where the red dashed line shows the two terms given in  Eq.~\eqref{eq:Approx-min-Var}.  For  $\alpha \to 1$, $\eta_\text{min}$ goes to zero as $\eta_\text{min} \propto (\alpha - 1)^{\frac{1}{2}}$. Recall that $\alpha > 1$  for a Gaussian process. 

The inset of Fig.~\ref{fig:Var-cut-num-1st} shows VBR$(\eta_{\rm min})/{\rm MBR}^2$, as a function of $\alpha$. We see that this ratio increases linearly with $\alpha$ for large $\alpha$, as expected from Eq.~\eqref{eq:Var_MBR_Gauss}, and asymptotically reaches zero for $\alpha \to 1$ with a power law of $(\alpha - 1)^\frac{1}{2}$. 

How does  $\alpha(\tau,t)$ depend on $\tau$ and $t$? To investigate this, we rewrite Eq.~\eqref{eq:a} purely in terms of MSD
\begin{align}
        \alpha(\tau,t) &
        = \frac{4\Delta x^2(t) \Delta x^2(\tau)}{\left( \Delta x^2(t+\tau) - \Delta x^2(t) - \Delta x^2(\tau) \right)^2}\notag\\
        &=\alpha(t,\tau),
\end{align}
i.e., $\alpha$ is symmetric in its arguments. In case of diffusive behavior, i.e.,  $\Delta x^2(t) \simeq 2Dt$ for large $t$, $\alpha(\tau,t)$ grows without bound with increasing time.
If MSD reaches a finite value, i.e., $ \displaystyle{\lim_{t \to \infty}}\Delta x^2(t) = A$, we obtain
\begin{align}
    \lim_{t \to \infty } \alpha(\tau,t) = \frac{4 A }{\Delta x^2(\tau)}.
\end{align}
If  MSD grows with time, there is thus a regime where $\alpha$ decreases with increasing $\tau$. With the symmetry $\alpha(\tau,t) = \alpha(t,\tau)$, the statement hols true for $\tau$ and $t$ exchanged. 

Summarizing, $\eta_{\rm min}$ takes values between zero and $0.65$, so that, especially, for large $\alpha$, $\eta\sim\mathcal{O}(1)$ is a useful rule of thumb, i.e., the length parameter that minimizes VBR may be estimated as of order $\sqrt{\Delta x^2(\tau)}$. VBR($\eta_{\rm min}$)/MBR$^2$ grows as a function of $\alpha$, which itself depends on $\tau$ and $t$. This analysis provides a guideline on how to choose the length parameter to evaluate MBR from (experimental) data.

\subsection{Variance: RHC model}
\label{sec:VBR-Cells-Model}

\begin{figure*}
    \centering
    \includegraphics[width=\linewidth]{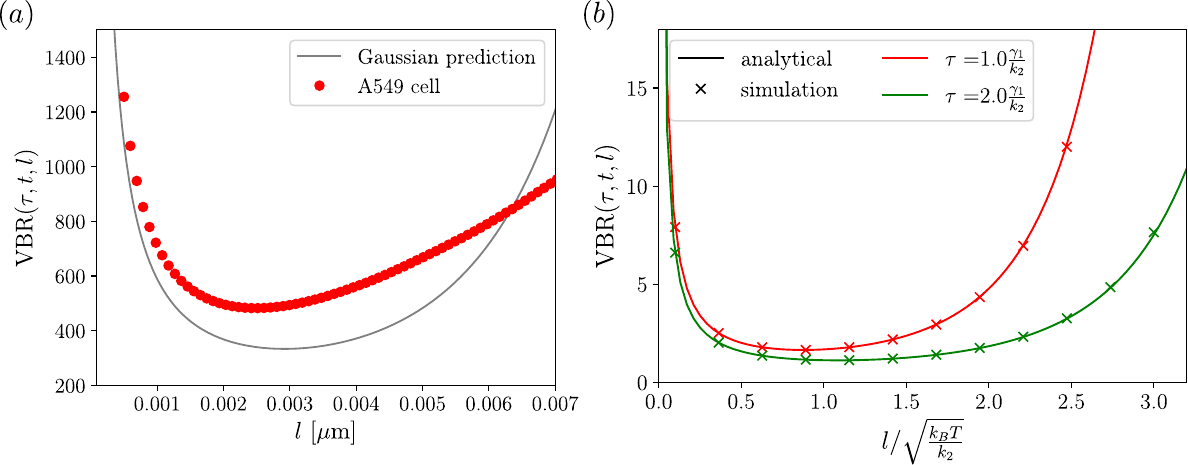}
    \caption{(a) VBR from A549 cells (red dots) compared to the Gaussian prediction using the particle's MSD via Eq.~\eqref{eq:Var_MBR_Gauss} (grey line) for $\tau = \SI{1}{\milli\second}$ and $t = \SI{1}{\second}$. Both curves show a similar qualitative behavior with a divergence for $l \to 0$ and $l \to \infty$. The two disagree, especially for larger $\tau$, demonstrating a non-Gaussianity of the process. (b) VBR of the RHC model of Eq.~\eqref{eq:model} against the length parameter $l$ for $\frac{k_1}{k_2} = 1.0, \frac{\gamma_1}{\gamma_2} = 0.2, \frac{D_q}{D_1} = 0.5$ and $t= 1.0 \frac{\gamma_1}{k_2}$ for two values of $\tau$ as labeled. The lines show the analytical prediction of Eq.~\eqref{eq:Var_MBR_Gauss} and crosses give numerical results from simulation. One can see the growth for low and high $l$. We note that the curve is rather flat around the minimum, which means that, especially for larger $\tau$, VBR is rather insensitive to the precise choice of $l$.}
    \label{fig:cell_HCM-cutoff}
\end{figure*}

The RHC model of Eq.~\eqref{eq:model} is Gaussian, so that VBR is given in Eq.~\eqref{eq:Var_MBR_Gauss}. Evaluating VBR for  RHC thus allows testing of Eq.~\eqref{eq:Var_MBR_Gauss} in numerical simulations \cite{rackauckas2017differentialequations} as well as estimating VBR for our experiments.  Figure~\ref{fig:cell_HCM-cutoff}(b) shows VBR as a function of the length parameter $l$, for fixed $t$ and $\tau$,  confirming Eq.~\eqref{eq:Var_MBR_Gauss}  using numerical simulations of RHC. We observe the mentioned divergence for small and large $l$. The amplitude of VBR is to a good estimate given by the ratio of  MSD evaluated at the times $t$ and $\tau$. Thus, increasing the time $\tau$ with keeping $t$ fixed leads to a reduction of VBR (This overdamped process has a monotonic MSD). Because the MSD increases linearly for large $t$, VBR, evaluated at $\eta_{\rm min}$ has no upper bound in this model. Therefore, the statistical error of MBR increases with $t$. Increasing $\tau$ on the other hand reduces VBR. Notably the curves are rather flat around the minimum, which softens the requirement of a precise choice of the length parameter $l$.

\subsection{Variance: Experiment}
For the data obtained in the mentioned cell, we find that VBR has a similar functional behavior as the Gaussian reference model. The dots show the variance directly obtained from the cell data, while the  line shows the prediction from MSD of this data using Eqs.~\eqref{eq:MBR-MSD} and \eqref{eq:Var_MBR_Gauss}. There is a surprising agreement between the two for  small $\tau$. However, one should keep in mind that  the experimental resolution is about 2nm, so that what is seen at smaller $l$ may be measurement noise. For larger values, the cell VBR grows and lies above the Gaussian prediction. From this, we conclude that the data obtained from cells  are notably non-Gaussian. 
However, since the curve is rather flat in the region of the minimum, the exact choice of the length parameter is not that important and the Gaussian estimate is still reasonable. Therefore, using the Gaussian estimate for the length parameter is also a good starting point for cell data.


\section{Discussion}
Mean back relaxation depends on two time parameters $\tau$ and $t$, and one length parameter $l$. Here we investigated in detail the dependence of MBR on $\tau$ and $t$, both in a linear driven RHC model system as well as for data obtained from cells.  The RHC model is found to qualitatively reproduce the characteristic shapes of MBR as a function of $t$ found in cells, namely a non-monotonic behavior with MBR first rising and then decaying with $t$. This shape as a function of $t$ is found in the investigated cells for any $\tau$, with however the height of the maximum as well as the final value strongly dependent on $\tau$. This behavior is also found in the RHC system for a certain regime of parameters. We find a strong dependence of MBR on $\tau$, which emphasizes the importance of this parameter. It is thus not only the future motion that is of interest (i.e., the dependence on $t$), but also the motion in the past, i.e.,  dependence on $\tau$.   
There are other parameter regimes where MBR, as a function of $t$ is monotonically decreasing or monotonicaly increasing. Notably, non-monotonic shapes cannot be found in the simpler model of Ref.~\onlinecite{muenker_accessing_2024}. The RHC model used here introduces one more (bath) particle, which yields memory, thus indicating that non-monotonic MBR behavior is a sign of memory.   

For large $\tau$, MBR decays to zero, found both in cell data, as well in the RHC model system. In the latter, it does so with a power law. Depending on the parameters, the power law can approach zero from above or below. 

The large $t$  limit of MBR shows, in cells, a minimum as a function of $\tau$. Such behavior is found in the RHC model system as well. The model furthermore shows  a $\tau$ dependent absolute minimum of MBR. This minimum goes to $-\infty$ for $\tau \to 0$, which suggest that for small $\tau$, MBR depends strongest on activity.  

This finding is of relevance when comparing MBR to effective energy. For  cell data, the effective energy shows a linear relation with the long time value of MBR. While this relation was observed in Ref.~\onlinecite{muenker_accessing_2024} for small values of $\tau$, we observe and establish it here for a range of $\tau$ values. For the RHC model system, this linear relation is also found, with however a limited range of validity, which decreases with increasing $\tau$. This can be attributed to the existence of the above mentioned absolute minimum of MBR. Future work has to investigate whether such limitations of the linear behavior can also be seen in experimental cellular systems.

For the Gaussian model studied, MBR does not depend on the length parameter $l$ \cite{knotz_mean_2024}, and, in absence of a model that allows discussion of $l$,  we have thus not investigated the $l$ dependence of MBR obtained in cells here. Investigation of the dependence on $l$ requires a non-Gaussian model, which we leave for future work. Indeed, it is an open question how the relation between MBR and effective energy depends on $l$. 

For the Gaussian model system, $l$ however appears in the variance of back relaxation (VBR), which we determine here for any Gaussian system. We find that VBR, as a function of $l$, shows a single minimum, which suggests that this value of $l$ yields the smallest statistical error bar of MBR. For cells, the theoretical prediction for VBR is observed qualitatively, with however some notable deviations, indicating that the cellular process is non-Gaussian. VBR therefore serves as a marker for non-Gaussianity.

Our investigations have revealed both striking similarities and important distinctions between the non-equilibrium Gaussian RHC model system and experimental observations in cells. The model successfully captures the qualitative behavior of MBR, and provides a framework for understanding the relationship between MBR and effective energy. However, there are also differences: The non-Gaussian character of the cellular process is visible in VBR. Furthermore, the dependence of effective energy on frequency is different, being  $\omega^{-2}$ and $\omega^{-1}$ for RHC model and cells, respectively. This suggests that there are limitations on how accurate the complex cellular dynamics can be approximated via the Gaussian process modeled here. An accurate description should include non-linear interactions, which, again, are left for future investigations.

The striking linear relation between MBR and effective energy requires more investigation, both experimental as well as theoretical. 

The linear relation between MBR and effective energy found in cell data for a larger range of $\tau$ values allows to  investigate this with lower temporal and spatial resolution setups, such as standard bright- or dark-field microscopes. 
The investigation of VBR suggests, as a rule of thumb,  to use $l \approx \sqrt{\Delta x^2(\tau)}$ in experimental applications. 

Future theoretical work can explore  potential connections between MBR and other non-equilibrium quantities, such as entropy production \cite{seifert_stochastic_2012,horowitz_thermodynamic_2020,roldan_quantifying_2021,knotz_entropy_2024}. It will also be interesting to evaluate MBR for other observables, such as density \cite{knotz_mean_2024}, to detect broken detailed balance in cells. The current study therefore provides valuable insights on promising parameter values. 

\begin{acknowledgments}
This work was supported by the German Research Foundation (DFG) under grant number KR 3844/5-1 (M.K.), and Project IDs 516046415; 456112451; 492390964 (T.B.).
\end{acknowledgments}

\section*{Author declarations}
\subsection*{Conflict of interest}
The authors have no conflicts to disclose.

\subsection*{Author Contributions}
\noindent \textbf{Gabriel Knotz}: Formal Analysis, Investigation, Software, Visualization, Writing - Original Draft. \\
\noindent \textbf{Till M. Muenker}: Formal Analysis, Investigation, Software, Visualization, Writing - Original Draft. \\
\noindent \textbf{Timo Betz}: Conceptualization, Methodology, Writing – Review \& Editing.\\
\noindent \textbf{Matthias Krüger}: Conceptualization, Methodology, Writing – Review \& Editing.

\section*{Data availability}
The data that support the findings of this study are available from the corresponding author upon reasonable request.

\bibliographystyle{ieeetr}
\bibliography{references}  
\appendix
\section{Methods}
\subsection{Variance of back relaxation}
\label{Methods:Var-MBR}
For purely Gaussian random variables, MBR only depends on the conditioning time $\tau$ and $t$, and is length parameter independent. The length parameter dependence enters when we consider VBR. First, we need to define the fluctuating back relaxation 
\begin{align}
    \mbr = -\frac{x(t) - x(0)}{x(0) - x(-\tau)} \vartheta_l(x(0) - x(-\tau)) = -\frac{b}{d}\vartheta_l(d)
\end{align}
with $\MBR(\tau,t,l) = \mean{\mbr}$. We assume that the joint process of $b$ and $d$ is also given by a Gaussian process. 
In this case the conditioned distribution is given by \cite{bishop_pattern_2006}
\begin{align}
    p(b\vert d) = \frac{1}{\sqrt{2 \pi \sigma^2}} e^{ - \frac{1}{2} \frac{\left( b + d\mean{\mbr} \right)^2}{\sigma^2}}
    \label{eq:condition}
\end{align}
with $\sigma^2 = \Delta x^2(t) - \Delta x^2(\tau) \frac{1}{4} \left(1 - \frac{\Delta x^2(t+\tau) - \Delta x^2(t)}{\Delta x^2(\tau)}\right) =  \Delta x^2(t) - \Delta x^2(\tau) \MBR^2(\tau,t)$, with the mean squared displacement (MSD) $\Delta x^2(t) = \mean{(x(t) - x(0))^2}$. Because the variance $\sigma^2$ needs to be positive, $\alpha = \frac{1}{\MBR^2(\tau,t)}\frac{\Delta x^2(t)}{\Delta x^2(\tau)} > 1$ . Note that in the Gaussian case MBR is independent of $l$. From here we proceed to calculate the distribution of the back relaxation 
\begin{align}
    p(\mbr) &= \int \dif d \dif b~ \delta \left( \mbr + \frac{b}{d} \vartheta_(d) \right) p(b\vert d) p(d)\\
    &= \int \dif d ~  p \left( \left. \frac{\mbr d}{\vartheta_l(d)}  \right\vert d \right) p(d) \frac{d}{\vartheta_l(d)}
\end{align}
Using Eq.~\eqref{eq:condition} and the Gaussian probability distribution 
\begin{align}
    p(d) = \frac{1}{\sqrt{2\pi \Delta x^2(\tau)}} e^{-\frac{1}{2} \frac{d^2}{\Delta x^2(\tau)}}
\end{align}
for the general displacements we can calculate VBR
\begin{widetext}
\begin{align}
    \VBR(\tau,t,l) &= \mean{\mbr^2} - \mean{\mbr}^2 \\
    &= \int \dif d \frac{\sigma^2 \vartheta_l^2(d)}{d^2} p(d) + \frac{\MBR^2(\tau,t)}{\text{erfc}\left(\frac{l}{\sqrt{2 \Delta x^2(\tau)}}\right)} - \MBR^2(\tau,t)\\
     &= \frac{\sigma^2}{\mean{\theta(\vert d \vert - l)}^2} \int \dif d \frac{\theta(\vert d \vert -l)}{d^2} \frac{1}{\sqrt{2 \pi \Delta x^2_\tau}} e^{ -\frac{d^2}{2 \Delta x^2_\tau}} + \MBR^2(\tau,t) \left( \frac{1}{\text{erfc}\left(\frac{l}{\sqrt{2 \Delta x^2(\tau)}}\right)} - 1\right)
\end{align}
\end{widetext}
where we inserted $\vartheta_l(d) = \frac{\theta(\vert d \vert  - l)}{\mean{\theta(\vert d \vert  - l)}}$. The remaining terms can be solved using the Gaussian error functions to get 
\begin{widetext}
\begin{align}
    \VBR(\tau,t,l) &=  \frac{\sigma^2}{\text{erfc} \left( \frac{l}{\sqrt{2 \Delta x^2(\tau)}} \right)^2} \left( \sqrt{\frac{2}{\pi}} \frac{e^{-\frac{l^2}{2\Delta x^2(\tau)}}}{\sqrt{\Delta x^2(\tau)}l} -\frac{1}{\Delta x^2(\tau)} \text{erfc} \left( \frac{l}{\sqrt{2 \Delta x^2(\tau)}} \right)\right)\\
    &+ \MBR^2(\tau,t) \left( \frac{1}{\text{erfc}\left(\frac{l}{\sqrt{2 \Delta x^2(\tau)}}\right)} - 1\right)\\
    &= \frac{\sigma^2}{\Delta x^2(\tau)} \frac{1}{ \text{erfc}(\eta)^2} \left( \frac{1}{\sqrt{\pi}} \frac{e^{-\eta^2}}{\eta} - \text{erfc}(\eta) \right) + \MBR^2(\tau,t) \left( \frac{1}{\text{erfc}(\eta)} - 1 \right)\\
    &= \left( \frac{\Delta x^2(t)}{\Delta x^2(\tau)} - \MBR^2(\tau,t) \right) \frac{1}{ \text{erfc}(\eta)^2} \left( \frac{1}{\sqrt{\pi}} \frac{e^{-\eta^2}}{\eta} - \text{erfc}(\eta) \right) +\MBR^2(\tau,t) \left(\frac{1}{\text{erfc}(\eta)} -1 \right)\\
    &= \MBR^2(\tau,t) \left[ \left( \alpha - 1\right) g(\eta) + h(\eta)\right]
\end{align}
\end{widetext}
with a dimensionless $\eta = \frac{l}{\sqrt{2 \Delta x^2(\tau)}}, \alpha = \frac{1}{\MBR^2(\tau,t)} \frac{\Delta x^2(t)}{\Delta x^2(\tau)}$ and the $\eta$ dependent functions $g(\eta)=\frac{1}{\text{erfc}(\eta)^2} \left( \frac{1}{\sqrt{\pi}} \frac{e^{-\eta^2}}{\eta} - \text{erfc}(\eta) \right)$ and $h(\eta) =  \frac{1}{\text{erfc}(\eta)} - 1$. Again note that $\MBR(\tau,t,l) = \mean{\mbr}$. The expression $g(\eta)$ can be minimized numerically with $\eta^{(g)}_\text{min} \approx 0.654334$. To get an approximation for the minimum of VBR up to first order in $y = \frac{1}{\alpha - 1}$ we express the derivative of VBR
\begin{align}
    \frac{\dif \VBR}{\dif \eta} &= \MBR^2 \left( \frac{1}{y} g^\prime(\eta_\text{min}) + h^\prime(\eta_\text{min}) \right) \overset{!}{=} 0\\
    \Rightarrow 0 &= g^\prime(\eta_\text{min}(y)) + y h^\prime(\eta_\text{min}(y))
\end{align}
Expand the last expression around $y=0$ to linear order and use $\eta(y) - \eta(0) = y\eta^\prime(0) + \mathcal{O}(y^2)$
\begin{align}
\begin{split}
0 =& g^\prime(\eta_\text{min}(0)) + g^{\prime\prime}(\eta_\text{min}(0)) (\eta_\text{min}(y) - \eta_\text{min}(0))\\
    &+ h^\prime(\eta_\text{min}(0)) y +\mathcal{O}(y^2).
    \end{split}
\end{align}
At last $\eta_\text{min}(0) = \eta_\text{min}^{(g)}$ and therefore $g^\prime(\eta_\text{min}(0)) = 0$
\begin{align}
    \eta_\text{min}(y) &= \eta_\text{min}^{(g)} - \frac{h^\prime \left(\eta_\text{min}^{(g)} \right)}{g^{\prime  \prime} \left(\eta_\text{min}^{(g)} \right)}y  + \mathcal{O}(y^2)\\
    &= \eta_\text{min}^{(g)} - \frac{h^\prime \left(\eta_\text{min}^{(g)} \right)}{g^{\prime  \prime} \left(\eta_\text{min}^{(g)} \right)} \frac{1}{\alpha - 1} + \mathcal{O} \left( \left( \alpha - 1 \right)^{-2} \right)
\end{align}

The expansion coefficients can be determined as $h^\prime(\hat{\eta}_\text{min}) = 5.84259$ and $g^{\prime \prime}(\hat{\eta}_\text{min}) = 6.71207$.

\subsection{Random Horse and Cart model}
\label{subsec:Methods:HCM}

\subsubsection{Mean back relaxation}~\\
\label{subsec:HCM-MBR}
The mean back relaxation (MBR) is related to the mean squared displacement $\Delta x^2(t) = \mean{(x(t) - x(0))^2}$ because the statistics of the overall system is Gaussian

 \begin{align}
     \MBR(\tau,t,l) = \frac{1}{2} \left(1 - \frac{\Delta x^2(t+\tau) - \Delta x^2(t)}{\Delta x^2(\tau)} \right)
 \end{align}
 which is independent of the length parameter $l$. We, therefore, drop this index for brevity.
 To get the mean squared displacement we rewrite the equations into relative coordinates $r_1 = x_1 - x_2$ and $r_2 = x_1 - q$
 \begin{subequations}
 \begin{align}
     \dot r_1 &= - k_1 \left( \frac{1}{\gamma_1} + \frac{1}{\gamma_2} \right) r_1 - \frac{k_2}{\gamma_1} r_2 + \xi_{r_1}(t)\\
     \dot r_2 &= - \frac{k_1}{\gamma_1} r_1 - \frac{k_2}{\gamma_1} r_2 + \xi_{r_2}(t)\\
     \dot q &= \xi_q(t)
 \end{align}
  \end{subequations}
  with $\xi_{r_1} = \xi_1(t) - \xi_2(t)$ and $\xi{r_2}(t) = \xi_1(t) - \xi_q(t)$
  The relative coordinates can be solved using the matrix equation 
  \begin{align}
      \mathbf{r}(t) = \int_{-\infty}^t e^{\mathbf{M} (t-s)} \mathbf{\xi_r}(s) \dif s
  \end{align}
  with 
  \begin{align}
      \mathbf{M} = \left( \begin{array}{cc}
           -k_1 \left( \frac{1}{\gamma_1} + \frac{1}{\gamma_2} \right)& - \frac{k_2}{\gamma_1}  \\
           -\frac{k_1}{\gamma_1}& -\frac{k_2}{\gamma_1} 
      \end{array}\right).
  \end{align}
  After diagonalizing $\mathbf{M}$ and computing the correlation functions, the mean squared displacement of the particle $x_1$ is found by $\Delta x_1^2(t) = \mean{(r_2(t) + q(t) - r_2(0) - q(0))^2}$
 \begin{widetext}
 \begin{align}
    \beta k_2 \Delta x_1^2(t) = 2 \frac{D_q}{D_1} \frac{k_2}{\gamma_1} t + \left(f(\lambda_1,\lambda_2) + g(\lambda_1,\lambda_2) \frac{D_q}{D_1} \right) \left(1 - e^{-\lambda_1 t} \right) + \left(f(\lambda_2,\lambda_1) + g(\lambda_2,\lambda_1) \frac{D_q}{D_1} \right) \left(1 - e^{-\lambda_2 t} \right)
    \label{eq:HCM-MSD_2}
\end{align}
 \end{widetext}
with time scales 
\begin{align}
\begin{split}
    \lambda_{1,2} = \frac{k_2}{\gamma_1} \frac{1}{2} &\left[ 1 + \frac{k_1}{k_2}+ \frac{\gamma_1}{\gamma_2} \frac{k_1}{k_2} \right.\\
    &\left. \pm \sqrt{ \left(1 + \frac{k_1}{k_2} + \frac{\gamma_1}{\gamma_2} \frac{k_1}{k_2} \right)^2 - 4 \frac{\gamma_1}{\gamma_2} \frac{k_1}{k_2} } \right]
    \end{split}
\end{align}
and constants
\begin{align}
    f(\lambda_1,\lambda_2) &= 2 \frac{\frac{k_2}{\gamma_1} - \lambda_2}{\lambda_1 - \lambda_2}\\
    g(\lambda_1,\lambda_2) &= 2 \frac{k_2}{\gamma_1} \frac{(\lambda_2 -\frac{k_2}{\gamma_1}) (\lambda_2 + \frac{k_2}{\gamma_1})}{(\lambda_1 + \lambda_2) (\lambda_1 - \lambda_2) \lambda_1}
\end{align}
By introducing the coefficients $\Psi(\lambda_1,\lambda_2) = f(\lambda_1,\lambda_2) + g(\lambda_1,\lambda_2) \frac{D_q}{D_1}$
\begin{widetext}
\begin{align}
    \MBR(\tau,t) = \frac{1}{2} \frac{\Psi(\lambda_1,\lambda_2) \left( 1 - e^{-\lambda_1 t} \right)\left( 1 - e^{-\lambda_1 \tau} \right) + \Psi(\lambda_2,\lambda_1) \left( 1 - e^{-\lambda_2 t} \right)\left( 1 - e^{-\lambda_2 \tau} \right)}{2 \frac{D_q}{D_1} \frac{k_2}{\gamma_1} \tau + \Psi(\lambda_1,\lambda_2) (1 - e^{-\lambda_1 \tau}) + \Psi(\lambda_2,\lambda_1) (1 - e^{-\lambda_2 \tau})}.
    \label{eq:MBR-HCM}
\end{align}
In the limit $\tau \to 0$ we obtain
\begin{align}
     \MBR(\tau \to 0,t) = \frac{1}{2} \frac{\lambda_1\Psi(\lambda_1,\lambda_2) \left(1 - e^{-\lambda_1 t} \right) + \lambda_2\Psi(\lambda_2,\lambda_1) \left(1 - e^{-\lambda_2 t} \right)}{2 \frac{D_q}{D_1} \frac{k_2}{\gamma_1}  +\lambda_1\Psi(\lambda_1,\lambda_2)  + \lambda_2\Psi(\lambda_2,\lambda_1) }
\end{align}
\end{widetext}
and in the long time limit $t \to \infty$ this expression simplifies to
\begin{align}
    \MBR(\tau \to 0,t \to \infty) = \frac{1}{2} \left( 1 - \frac{D_q}{D_1}\right)
\end{align}
\subsubsection{Maximum of MBR}~\\
The MBR has for certain parmeters $\frac{\gamma_1}{\gamma_2}, \frac{k_1}{k_2}$ a maximum in $t$ which one can obtain by setting the derivative of Eq.\eqref{eq:MBR-HCM} to zero. The result is given by
\begin{align}
    t_\text{max} = \frac{1}{\lambda_1 - \lambda_2} \log \left( - \frac{\lambda_1 \Psi(\lambda_1, \lambda_2) \left(1 - e^{-\lambda_1 \tau} \right)}{\lambda_2 \Psi(\lambda_2, \lambda_1) \left(1 - e^{-\lambda_2 \tau} \right)} \right).
\end{align}
However, not for every time scales $\lambda_1,\lambda_2$ there is a finite time solution as is illustrated in Fig.~\ref{fig:HCM-MBR-typical} because the expression in the log can be negative or smaller than 1. For activities smaller than 
\begin{align}
    \frac{D_q}{D_1} < \frac{(\lambda_1 + \lambda_2)\lambda_2}{\frac{k_2}{\gamma_1}+ \lambda_1}
\end{align}
there is no maximum, which is remarkably independent of $\tau$. Also for too high activities there is no maximum, which we find by setting $\displaystyle{\lim_{t \to 0}} \frac{\dif}{\dif t} \MBR(\tau,t) \overset{!}{=} 0$
\begin{align}
\begin{split}
    &\frac{D_q}{D_1}\\ &>- \frac{\lambda_1 f(\lambda_1,\lambda_2) \left(1 - e^{-\lambda_1 \tau} \right) + \lambda_2 f(\lambda_2,\lambda_1) \left(1 - e^{-\lambda_2 \tau} \right)}{\lambda_1 g(\lambda_1,\lambda_2) \left(1 - e^{-\lambda_1 \tau} \right) + \lambda_2 g(\lambda_2,\lambda_1) \left(1 - e^{-\lambda_2 \tau} \right)},
    \end{split}
\end{align}
however, this bound is $\tau$ dependent. In the limit $\tau \to 0$ we obtain Eq.~\eqref{eq:cond3}.
\subsubsection{Critical Activity}~\\
The critical activity is given by the ration of active to passive diffusion $\frac{D_q}{D}$, for which the long time MBR gets negative and has a local minimum. This property depends on the timescales of the system $\lambda_1$ and $\lambda_2$. The long time MBR for $\tau \gg \lambda_1,\lambda_2$ is given by
\begin{align}
    \MBR(\tau,t\to \infty) = \frac{1}{2} \frac{\Psi(\lambda_1,\lambda_2) + \Psi(\lambda_2,\lambda_1)}{2 \frac{D_q}{D_1} \tau + \Psi(\lambda_1,\lambda_2) + \Psi(\lambda_2,\lambda_1)}.
\end{align}
Because $\frac{D_q}{D_1} > 0$ the sign of the long time MBR is determined by the sign of $\Psi(\lambda_1,\lambda_2)+\Psi(\lambda_2,\lambda_1)$. Using $\Psi(\lambda_1,\lambda_2) = f(\lambda_1,\lambda_2) + \frac{D_q}{D_1} g(\lambda_1,\lambda_2)$ we find for the critical activity  $\Psi(\lambda_1,\lambda_2) + \Psi(\lambda_2,\lambda_1) = 0$
\begin{align}
    \frac{D_q}{D} = -\frac{f(\lambda_1,\lambda_2)  + f(\lambda_2,\lambda_1) }{g(\lambda_1,\lambda_2)  + g(\lambda_2,\lambda_1) }
    \label{eq:HCM-act-crit2}
    \end{align}
\subsubsection{Effective energy}~\\
The diffusing potential with memory can be described by the following Langevin equations

\begin{subequations}
\begin{align}
    \dot x_1 &= -\frac{k_1}{\gamma_1} (x_1 - x_2) - \frac{k_2}{\gamma_1} (x_1 - q) + \xi_1(t)\\
    \dot x_2 &= - \frac{k_1}{\gamma_2} (x_2 - x_1) + \xi_2(t)\\
    \dot q &= \xi_q
\end{align}
\end{subequations}

with $x_1$ the tracer particle, $x_2$ the bath particle that mimics the viscoelastic environment and $q$ the trap position. The system is projected onto $x_1$ coordinate and by integrating out the bath particle a generalized Langevin equation can be found by
\begin{align}
    -k_2x(t) + \int_{-\infty}^t \Gamma(t-s) \dot x(s) = k_2q(t) + f(t)
\end{align}
with a Gaussian random force $\mean{f(t) f(t^\prime)} = k_B T \Gamma (\vert t- t^\prime \vert)$. The correlation function in Fourier space is therefore given by
\begin{align}
    \mean{x(\omega) x(-\omega)} = \frac{k_2^2 \mean{q(\omega) q(-\omega) + \mean{f(\omega)f(-\omega)}}}{\omega^2 \Gamma(\omega) \Gamma(-\omega) + k_2^2 + ik_2\omega (\Gamma(\omega) - \Gamma(-\omega))} 
\end{align}
To get the effective energy we compare the correlation function with the active driving $q$ to the correlation function without the active driving and obtain
\begin{align}
    \frac{E_\text{eff}(\omega)}{k_B T} &= 1 + \frac{k_2^2 \mean{q(\omega)q(-\omega)}}{\mean{f(\omega) f(-\omega)}}\\
    &= 1 + \frac{k_2^2 \mean{q(\omega)q(-\omega)}}{k_B T \Tilde{\Gamma} (\omega)}\\
    &= 1 + \frac{D_q}{D} \frac{k_2^2}{\gamma_1} \frac{1}{\omega^2}\frac{1}{\Tilde{\Gamma} (\omega)}
\end{align}
with $\Tilde{\Gamma}(\omega)$ the Fourier transform of $\Gamma(\vert t\vert)$ and the Fourier transform of the free diffusion $\mean{q(\omega) q(-\omega)} = \frac{D_q}{\omega^2}$ with $D = \frac{k_B T}{\gamma_1}$. For the model system the memory kernel is given by
\begin{align}
    \Gamma(t) = \left[2 \gamma_1 \delta(t) + k_1 e^{-\frac{k_1}{\gamma_2} t} \right] \theta(t)
\end{align}
with the Fourier transform $\Tilde{\Gamma}(\omega) = \gamma_1 + \frac{k_1^2}{\gamma_2} \frac{1}{\frac{k_1^2}{\gamma_2^2} + \omega^2 }$ we get the final expression for the effective energy in the main text that scales with $\frac{D_q}{D}$.
    
\subsection{Active and passive microrheology}
\label{Methods: Active and passive microrheology}
The experiments conducted in order to determine the complex shear modulus and the effective energy of different cell types have been described in detail elsewhere \cite{muenker_accessing_2024}. Briefly, a home-build optical tweezers setup with a movable \SI{808}{\nano\meter} laser (LU0808M250, Lumics, \SI{75}{\milli\watt} at sample plane) and an \SI{976}{\nano\meter} laser (BL976-PAG500, Throlabs, \SI{0.3}{\milli\watt} at sample plane) were focused by a high NA objective (CFI Plan Apochromat VC 60XC WI NA 1.2, Nikon) onto the sample to exert well defined forces and detect particle motion.\\ \newline
For active microrheology, the \SI{808}{\nano\meter} trapping laser was used to apply oscillatory forces at different frequencies ($f_{\text{Drive}} \in \{1,2,4,8,16,32,64,128,256,512,1025\} \si{\hertz}$) onto phagocytosed probe particles in the cytoplasm of cells. The resulting forces $F_{f_{\text{Drive}}}(t)$ were directly obtained using back focal plane interferometry \cite{farre2010force}.
Simultaneously, the second \SI{976}{\nano\meter} laser (detection laser) was directed at the probe particle to monitor its displacement $x_{f_{\text{Drive}}}(t)$ in response to the applied force.\\
For each driving frequency, the response function $\chi({f_{\text{Drive}}})$ was calculated according to 
\begin{align}
    \chi({f_{\text{Drive}}})= \frac{\Tilde{x}_{f_{\text{Drive}}}(t)}{\Tilde{F}_{f_{\text{Drive}}}(t)},
\end{align}
where the tilde denotes the Fourier Transformation.\\
\newline
For passive microrheology, only the \SI{976}{\nano\meter} detection laser was used to monitor the free particle fluctuations $x(t)$ in the absence of any relevant trapping forces. These passive trajectories were later used to calculate MBR. Additionally, $x(t)$ was used to calculate the power spectral density $C(f)$:
\begin{align}
    C(f)=\mean{\Tilde{x}(f)\Tilde{x}(f)^*},
\end{align}
with $\Tilde{x}(f)^*$ denoting the complex conjugate of the Fourier Transformation of $x(f)$.\\
\newline
The effective energy $E_\text{Eff}(f)$ was determined by combining results from passive and active microrheology:
\begin{align}
    E_\text{Eff}(f) = \frac{C(f_{\text{Drive}})\pi f_{\text{Drive}}}{ \chi^{\prime\prime}({f_{\text{Drive}}})}.
\end{align}
Finally, effective energies were fit with 
\begin{align}
    E_{\text{Eff}}(f)=E_0 \left( \frac{f_0}{f} \right)^{\nu}+ k_B T ,
\end{align}
with $f_0 = \SI{1}{\hertz}$ in order to obtain the effective energy amplitude $E_0$. In this publication, the average $E_0$ values for each cell type and condition are displayed.
\subsection{Biological samples}
\label{Methods: Cells}
The exact procedure of cell preparation has been described elsewhere in detail \cite{muenker_accessing_2024}. Briefly, A549 (N=61), C2C12 (N=59), CT26 (N=60), HeLa (N=79), MCF7 (N=60), MDCK (N=62) and HoxB8 (N=60) cells were seeded onto a coverslip previously incubated with 1\% Fibronectin diluted in PBS. After the cells attached, the medium was replaced with fresh medium containing probe particles (Polybead® Microspheres $\SI{1}{\micro \metre}$) at 1:10,000 dilution. After approximately 17 hours of incubation, the samples were washed with PBS. During this time, a portion of probe particles got phagocytosed by the cells. Then a second cover slip was placed over the sample with a $\SI{400}{\micro\metre}$ thick, double-sided adhesive strip as a spacer. After this step, the active and passive microrheology experiments were conducted.\\

\subsection{MBR calculation}
\label{Methods: MBR calculation}
MBR was calculated for trajectories $x_i=x(t_i)$ with positions recorded at discrete time points $t_i$, with a sampling rate $srt$ and being $N$ samples long. We calculated MBR for different lagtimes $t$, length parameter values $l$ and conditioning times $\tau$: $\text{MBR}(t,l,\tau)$. We chose each time-point $t_i$ of the trajectory as a starting point for the calculation of the back relaxation and then took the average to calculate the mean back relaxation. The back relaxation for a position point at time $t_i$ is given as ,
\begin{align}
\begin{split}
    \mbr(t,\tau,l_i)_i&= -\frac{x(t_i+t)-x(t_i)}{x(t_i)-x(t_i-\tau)}\\
    &\text{ if } (l_i = x(t_i)-x(t_i-\tau) > l) 
    \end{split}
\end{align}
but we only consider starting times $t_i$ were $l_i$ is larger than the length parameter $l$.\\
Then we calculated MBR as the average overall $N$ back relaxations where $l_i >=l$:
\begin{align}
    \text{MBR}(t) = \frac{\sum_{i=1}^{N} \mbr(t, \tau, l_i) \text{ if } (l_i > l)}{\sum_{i=1}^{N} 1 \text{ if } (l_i > l)}
\end{align}

\subsection{Statistical tests}
\label{Methods: Statistical tests}
\subsubsection{Determination of $R^2$-values}
\label{Methods: R2 values calculation}
The coefficient of determination, $R^2$, was calculated to assess the goodness of fit for the regression models. First, the independent variable values $x$ and the dependent variable values $y$ were provided, along with the model function used for fitting the data. Optimal values for the parameters of the model function were obtained from curve fitting. For this, the \textit{scipy} function \textit{curve\_fit} was used \cite{2020SciPy-NMeth}. The model function was then evaluated at each value in $x$ using these optimal parameters to obtain the fitted values $ y_{fit}$.\\
The residual sum of squares $SS_{res}$ was computed as the sum of the squared differences between the observed values $y$ and the fitted values $y_{fit}$:
\begin{align}
    SS_{res} = \sum (y_{fit} - y)^2
\end{align}
Next, the total sum of squares $SS_{tot}$ was calculated as the sum of the squared differences between the observed values $y$ and their mean value $\bar{y}$:
\begin{align}
    SS_{tot} = \sum (y - \bar{y})^2
\end{align}
The $R^2$ value was then derived, indicating the proportion of the variance in the dependent variable that is predictable from the independent variables:
\begin{align}
    R^2 = 1 - \frac{SS_{res}}{SS_{tot}}
\end{align}

\section*{Supplementary Information}

\subsection*{Supplementary Tables}
\begin{table}[H]
\centering
\caption{$R^2$-values of fits of $E_0$ to $\text{MBR}(t\to\infty)$ relation for different $\tau$-values}
\label{si: r2 values e0 mbr fit}
\begin{tabular}{ll}
$\tau [s]$ & $R^2$-value \\ \hline
5e-05  & 0.9333      \\
0.0001 & 0.96        \\
0.0002 & 0.9493      \\
0.0004 & 0.9449      \\
0.0006 & 0.9636      \\
0.0008 & 0.977       \\
0.001  & 0.9658      \\
0.002  & 0.9702      \\
0.003  & 0.9891      \\
0.004  & 0.9929      \\
0.007  & 0.9911      \\
0.008  & 0.9935      \\
0.01   & 0.993      
\end{tabular}
\end{table}

\end{document}